\documentclass[conference]{IEEEtran}
\pdfoutput=1
\usepackage{float}
\usepackage{amsfonts}
\usepackage{amsthm}
\usepackage{amsmath}
\usepackage{amssymb}
\usepackage{graphicx}
\usepackage{hyperref}
\usepackage{algorithm}
\usepackage{algpseudocode}
\floatstyle{ruled}
\newfloat{algorithm}{tbp}{loa}
\providecommand{\algorithmname}{Algorithm}
\floatname{algorithm}{\protect\algorithmname}

\DeclareMathOperator{\supp}{supp}
\DeclareMathOperator{\sgn}{sgn}
\newcommand{\GF}{\mathrm{GF}}
\newcommand{\llrbox}{\mathbin{\scalebox{0.75}{$\square$}}}
\newcommand{\starbox}{\mathbin{\setbox0\hbox{$\square$}
\setbox1\hbox to \wd0{\hss$\star$\hss}
\protect\rlap{\box0}{\protect\raisebox{0.25ex}{\box1}}
}}

\begin{document}
\title{Construction and analysis of polar and concatenated polar codes: 
practical approach}
\author{\authorblockN{Gregory Bonik}
\authorblockA{Dept. of Mathematics\\
 University of Connecticut\\
 196 Auditorium Road, Unit 3009\\
Storrs, CT 06269-3009, USA\\
Email: grigory.bonik@uconn.edu}
\and
\authorblockN{Sergei Goreinov}
\authorblockA{Institute of Numerical\\
Mathematics, R.A.S.\\
Gubkina 8, 119333 Moscow, Russia\\
Email: sergei@inm.ras.ru}
\and
\authorblockN{Nickolai Zamarashkin}
\authorblockA{Institute of Numerical\\
Mathematics, R.A.S.\\
Gubkina 8, 119333 Moscow, Russia\\
Email: kolya@bach.inm.ras.ru}}
\maketitle
\begin{abstract}
We consider two problems related to polar codes. First is the problem of polar codes construction and analysis
of their performance without Monte-Carlo method. The formulas proposed
are the same as those in [Mori-Tanaka], yet we believe that our approach is original and has clear
advantages. The resulting computational procedure is presented in a fast algorithm form which can be easily implemented
on a computer.
Secondly, we present an original method of construction of concatenated codes based on polar codes. We give an algorithm
for construction of such codes and present numerical experiments showing significant performance improvement with respect to
original polar codes proposed by Ar\i kan. We use the term \emph{concatenated code} not in its classical
sense (e.g. [Forney]). However we believe that our usage is quite appropriate for the exploited construction.
Further, we solve the optimization problem of choosing codes minimizing the block
error of the whole concatenated code under the constraint of its fixed rate.
\end{abstract}
\section{Introduction}

Research related to construction of coding systems whose performance is close to Shannon limit while
encoding and decoding algorithms are of low complexity, goes for more than 60 years.

A significant modern example of such system is linear codes with low-density parity checks (LDPC).
Usually these are binary linear block codes with sparse parity-check matrix. Decoding is performed
via iterative algorithms whose convergence is described by quite a few theoretical results
and in general, these algorithms are quite good.
In practice, LDPC codes have good performance for high noise levels.
There exist experimentally constructed codes approaching Shannon limit very closely (e.g. \cite{Chung01onthe}).
However for high bandwidth region, short LDPC codes exhibit so-called ``error floor'', i.e. significant slowdown
in decrease of decoding error probability corresponding to channel improvement, occurring due to decoding algorithm features.

Polar codes were invented by E.~Ar\i kan in 2008. They are the first coding system possessing, on the theorem level,
the convergence to Shannon limit for code length $N\to\infty$, as well as fast encoding/decoding algorithms with complexity bound
$O(N \log_2 N)$. Thus polar codes are a significant theoretical result.

On the other hand, the performance of polar codes in their initial form presented by Ar\i kan, is considerably inferior,
for a fixed code length, to other coding systems. To date there exist many proposals for improvement of polar codes performance
(e.g.  \cite{DBLP:journals/corr/abs-1001-2545,trif1}), yet work in this direction seems to be very promising.

In this paper we consider two problems related to polar codes. First is the problem of polar codes construction and analysis
of their performance for various types of binary symmetric channel without Monte Carlo method. The formulas proposed
are the same as those in \cite{DBLP:journals/corr/abs-0901-2207}, yet we believe that our approach is original and has clear
advantages. Moreover, the resulting computational procedure is presented in a fast algorithm form which can be easily implemented
on a computer.
Secondly, we present an original method of construction of concatenated codes based on polar codes. We give an algorithm
for construction of such codes and present numerical experiments showing significant performance improvement with respect to
original polar codes proposed by Ar\i kan. It should be noted that we use the term \emph{concatenated code} not in its classical
sense (e.g. \cite{Forney-1}). However we believe that our usage is quite appropriate for the exploited construction.
Our idea is simple. It is known that approaching the Shannon limit is possible only with sufficiently large code length.
Increasing the code length however makes the problem of code construction with large minimum distance and efficient ML decoder
very hard. The situation is different for low-noise channel. Here codes of moderate length are sufficient
so that ML decoder complexity is not too large. In order to obtain those low-noise channels we employ the polarization effect
observed by E.~Ar\i kan in polar codes. Further, we solve the optimization problem of choosing codes minimizing the block
error of the whole concatenated code under the constraint of its fixed rate.

Unfortunately, we do not have a theorem on asymptotic optimality of our approach or just on its clear advantage with respect
to known approaches, like e.g.  \cite{DBLP:journals/corr/abs-1001-2545}. Yet the simplicity of our approach, its
flexibility and further possibilities of its improvement make it hopefully interesting.

Other examples of concatenated and generalized concatenated codes based on polar codes can be found in e.g.
\cite{DBLP:journals/corr/abs-1001-2545,trif1}. 

A word on the channels considered here.
We assume that the channel is defined by input alphabet
$\mathcal{X}$, output alphabet
$\mathcal{Y}$ and transition function
\[
W(y\,|\, x):\quad\mathcal{Y}\times\mathcal{X}\to[0,1]
\]
defined for any pair $x\in\mathcal{X},\, y\in\mathcal{Y}.$
The function $W(y\,|\, x)$ defines the probability (or its density)
that symbol $y$ is received under the condition that symbol $x$ was sent.
For the sake of simplicity and in order to avoid generalized distributions
we restrict our discussion to finite output alphabet. All formulas can be easily
transplanted to the case of continuous channel by replacing the probabilities
by the probability densities and replacing some sums by integrals. Note also
that most frequently used channel models can be approximated by discrete ones.
Moreover, data transmission systems used in practice represent output symbols
with some fixed accuracy which is equivalent to some discrete channel model.

Besides, we consider only symmetric channels with binary input
\cite{Richardson:2008:MCT:1795974}. In such channels, the input alphabet contains
two symbols:
\[
\mathcal{X}=\{0,1\}=\GF(2),
\]
Output alphabet $\mathcal{Y}$ is a subset of real numbers, and the function
$W(y\,|\, x)$ possesses the following symmetry,
\[
W(y\,|\,0)=W(-y\,|\,1).
\]

The rest of this paper contains 5 sections. In section~\ref{sec:bp-on-trees}
we consider the problem of obtaining the optimal statistical estimate of a bit
variable restricted by a linear system. Results of this auxiliary section are well known
and belong to factor graph theory. These results are used in obtaining relations which
determine the probability of erroneous bit decoding for ML decoder in section~\ref{sec:polar}.
Our derivation essentially differs from original one proposed by E.~Ar\i kan.
It is based on explicit representation of factor graph of polar code, its interpretation
as a set of trees and application of density evolution method.
Note also that for polar codes we consider two types of factor graphs: encoder graph and decoder graph.
In section~\ref{sec:design} we describe the polar code construction method in the form of
fast algorithms taking on input discrete probability function defined by the channel.
Presented also is the analysis of obtained codes and numerical simulation for polar codes of different
length.

In section~\ref{sec:kernels} we discuss the possibility of polar code construction using polarization
kernels other than $G_2$ which was introduced in Ar\i kan's paper.
Finally, in section~\ref{sec:new-polar} we introduce a class of concatenated codes
based on polar codes and present numerical comparison of concatenated and classical polar codes performance.

\section{Problem of bit variable estimation\label{sec:bp-on-trees}}

Before proceeding directly to polar codes, consider the problem of estimation of one random bit
entering as a variable in a linear system. To this end we investigate two simpler problems:
estimation of sum of two random bits transmitted through the channels and estimation of one random
bit for which we have several independent sources of information.
Actually, this section contains short presentation of factor graph theory which is widely
used in the modern coding theory \cite{Richardson:2008:MCT:1795974}.

\subsection{Estimation of sum of bits\label{sub:check-node-update}}

Let the values of two independent random bits $x_1$ and $x_2$
taking the values $0$ and $1$ equiprobably, were transmitted through channels $W_1$ and $W_2$,
respectively, which resulted in received symbols $y=[y_{1},y_{2}]$.
Using the channel model we compute the logarithmic likelihood ratios (LLRs)
\[
L(x_{i})=\ln l(x_{i})=\ln\frac{\Pr\big\{ y_{i}\,|\, x_{i}=0\big\}}{\Pr\big\{ y_{i}\,|\, x_{i}=1\big\}}=\ln\frac{W_{i}(y_{i}|0)}{W_{i}(y_{i}|1)},\quad i=1,2.
\]
Assume the following quantity is required
\[
L(x_{1}\oplus x_{2})=\ln l(x_{1}\oplus x_{2})=\ln\frac{\Pr\big\{ y\,|\, x_{1}\oplus x_{2}=0\big\}}{\Pr\big\{ y\,|\, x_{1}\oplus x_{2}=1\big\}},
\]
i.e. estimate the sum of two bits provided $L(x_1)$ and $L(x_2)$ are known.
Considering two possible equiprobable cases, we get
\begin{eqnarray*}
\Pr\Big\{ y\,\big|\, x_{1}\oplus x_{2}=0\Big\}&=&\frac{1}{2}\Pr\Big\{ y\,\big|\, x_{1}=0,x_{2}=0\Big\}\\
&+&\frac{1}{2}\Pr\Big\{ y\,\big|\, x_{1}=1,x_{2}=1\Big\}.
\end{eqnarray*}
Since the bits $y_1$ and $y_2$ are transmitted independently,
\begin{eqnarray*}
\Pr\Big\{ y\,\big|\, x_{1}=0,x_{2}=0\Big\} & = & \Pr\Big\{ y_{1}\,\big|\, x_{1}=0\Big\}\Pr\Big\{ y_{2}\,\big|\, x_{2}=0\Big\},\\
\Pr\Big\{ y\,\big|\, x_{1}=1,x_{2}=1\Big\} & = & \Pr\Big\{ y_{1}\,\big|\, x_{1}=1\Big\}\Pr\Big\{ y_{2}\,\big|\, x_{2}=1\Big\}.
\end{eqnarray*}
Hence
\begin{eqnarray*}
\Pr\Big\{ y\,\big|\, x_{1}\oplus x_{2}=0\Big\} & = & \frac{1}{2}\Pr\Big\{ y_{1}\,\big|\, x_{1}=0\Big\}\Pr\Big\{ y_{2}\,\big|\, x_{2}=0\Big\}\\
 & + & \frac{1}{2}\Pr\Big\{ y_{1}\,\big|\, x_{1}=1\Big\}\Pr\Big\{ y_{2}\,\big|\, x_{2}=1\Big\}.
\end{eqnarray*}
In a similar way we get
\begin{eqnarray*}
\Pr\Big\{ y\,\big|\, x_{1}\oplus x_{2}=1\Big\} & = & \frac{1}{2}\Pr\Big\{ y_{1}\,\big|\, x_{1}=0\Big\}\Pr\Big\{ y_{2}\,\big|\, x_{2}=1\Big\}\\
 & + & \frac{1}{2}\Pr\Big\{ y_{1}\,\big|\, x_{1}=1\Big\}\Pr\Big\{ y_{2}\,\big|\, x_{2}=0\Big\}.
\end{eqnarray*}
Inserting the last two formulas in likelihood ratio
$l(x_{1}\oplus x_{2})$ and cancelling the factor $\frac{1}{2}$, we obtain
\begin{eqnarray*}
\quad l(x_{1}\oplus x_{2})=\hskip20em\\
\frac{\Pr\{y_{1}\,|\, x_{1}=0\}\Pr\{y_{2}\,|\, x_{2}=0\}+\Pr\{y_{1}\,|\, x_{1}=1\}\Pr\{y_{2}\,|\, x_{2}=1\}}{\Pr\{y_{1}\,|\, x_{1}=0\}\Pr\{y_{2}\,|\, x_{2}=1\}+\Pr\{y_{1}\,|\, x_{1}=1\}\Pr\{y_{2}\,|\, x_{2}=0\}}.\\
\end{eqnarray*}
Divide the numerator and the denominator by $\Pr\{y_{1}\,|\, x_{1}=1\}$ $\Pr\{y_{2}\,|\, x_{2}=1\}$:
\[
l(x_{1}\oplus x_{2})=\frac{\frac{\Pr\big\{ y_{1}\,|\, x_{1}=0\big\}}{\Pr\big\{ y_{1}\,|\, x_{1}=1\big\}}\cdot\frac{\Pr\big\{ y_{1}\,|\, x_{2}=0\big\}}{\Pr\big\{ y_{1}\,|\, x_{2}=1\big\}}+1}{\frac{\Pr\big\{ y_{1}\,|\, x_{1}=0\big\}}{\Pr\big\{ y_{1}\,|\, x_{1}=1\big\}}+\frac{\Pr\big\{ y_{1}\,|\, x_{2}=0\big\}}{\Pr\big\{ y_{1}\,|\, x_{2}=1\big\}}}.
\]
Using the likelihood ratios $l(x_{1})$ and $l(x_{2})$, rewrite the last formula as follows,
\[
l(x_{1}\oplus x_{2})=\frac{l(x_{1})l(x_{2})+1}{l(x_{1})+l(x_{2})},
\]
or using logarithms,
\begin{eqnarray*}
L(x_{1}\oplus x_{2})&=&\ln\frac{e^{L(x_{1})+L(x_{2})}+1}{e^{L(x_{1})}+e^{L(x_{2})}}\\
&=&2\tanh^{-1}\left(\tanh\left(\frac{L(x_{1})}{2}\right)\tanh\left(\frac{L(x_{2})}{2}\right)\right).
\end{eqnarray*}
For convenience introduce the binary operation
\[
a\llrbox b\equiv2\tanh^{-1}\left(\tanh\Big(\frac{a}{2}\Big)\tanh\Big(\frac{b}{2}\Big)\right).
\]
Now,
\[
L(x_{1}\oplus x_{2})=L(x_{1})\llrbox L(x_{2}).
\]
Note some useful properties of the $\llrbox$ operation:
\begin{eqnarray*}
a\llrbox b & = & b\llrbox a,\quad\forall a,b\in\overline{\mathbb{R}}\\
a\llrbox(b\llrbox c) & = & (a\llrbox b)\llrbox c,\quad\forall a,b,c\in\overline{\mathbb{R}},\\
a\llrbox0 & = & 0,\quad\forall a\in\overline{\mathbb{R}},\\
(-a)\llrbox b & = & -(a\llrbox b),\quad\forall a,b\in\overline{\mathbb{R}},\\
a\llrbox+\infty & = & a,\quad\forall a\in\overline{\mathbb{R}},\\
a\llrbox-\infty & = & -a,\quad\forall a\in\overline{\mathbb{R}},\\
|a\llrbox b| & \leq & \min(|a|,|b|),\quad\forall a,b\in\overline{\mathbb{R}}\\
\sgn(a\llrbox b) & = & \sgn a\cdot\sgn b,\quad\forall a,b\in\overline{\mathbb{R}}.
\end{eqnarray*}
We now extend the problem to three bits $x_{1},x_{2},x_{3}.$ 
Let these quantities be transmitted via channels $W_{1},W_{2},W_{3}$, respectively,
and symbols  $y=[y_{1},y_{2},y_{3}]$ be received. Assume
the following quantity is required
\begin{eqnarray*}
L(x_{1}\oplus x_{2}\oplus x_{3})&=&\ln l(x_{1}\oplus x_{2}\oplus x_{3})\\
&=&\ln\frac{\Pr\big\{ y\,|\, x_{1}\oplus x_{2}\oplus x_{3}=0\big\}}{\Pr\big\{ y\,|\, x_{1}\oplus x_{2}\oplus x_{3}=1\big\}}.
\end{eqnarray*}
Introduce new variable $t$ taking values $0$ and $1$ equiprobably:
\[
t=x_{1}\oplus x_{2}.
\]
We can assume that $t$ was transmitted via channel with the following transition function,
\[
W(y_{1}y_{2}\,|\, t)=\Pr\Big\{ y_{1}y_{2}\,|\, x_{1}\oplus x_{2}=t\Big\},
\]
and write its LLR value as
\begin{eqnarray*}
L(t)&=&\ln\frac{\Pr\Big\{ y_{1}y_{2}\,|\, x_{1}\oplus x_{2}=0\Big\}}
               {\Pr\Big\{ y_{1}y_{2}\,|\, x_{1}\oplus x_{2}=1\Big\}}\\
&=&L(x_{1}\oplus x_{2})=L(x_{1})\llrbox L(x_{2}).
\end{eqnarray*}
Then
\begin{eqnarray*}
L(x_{1}\oplus x_{2}\oplus x_{3})&=&L(t\oplus x_{3})=L(t)\llrbox L(x_{3})\\
&=&(L(x_{1})\llrbox L(x_{2}))\llrbox L(x_{3}).
\end{eqnarray*}
Since the $\llrbox$ operation is associative, drop the parentheses:
\[
L(x_{1}\oplus x_{2}\oplus x_{3})=L(x_{1})\llrbox L(x_{2})\llrbox L(x_{3}).
\]
Using induction, we obtain formula for arbitrary number of variables:
\[
L(x_{1}\oplus x_{2}\oplus\ldots\oplus x_{n})=L(x_{1})\llrbox L(x_{2})\llrbox\ldots\llrbox L(x_{n}).
\]
We now proceed to estimation of bit transmitted independently via several channels.

\subsection{Estimation of bit transmitted several times}

Let random bit $x$ taking values $0$ and $1$ equiprobably be transmitted via $n$
different channels
$W_{1},\ldots,W_{n},$ receiving symbols $y=[y_{1},\ldots,y_{n}].$
One can compute LLRs relying only on one channel:
\[
L_{i}(x)=\ln l_{i}(x)=\ln\frac{\Pr\big\{ y_{i}\,|\, x=0\big\}}{\Pr\big\{ y_{i}\,|\, x=1\big\}}=\ln\frac{W_{i}(y_{i}|0)}{W_{i}(y_{i}|1)},\quad i=\overline{1,n}.
\]
We need to estimate $x$, that is
\[
L(x)=\ln l(x)=\ln\frac{\Pr\big\{ y\,|\, x=0\big\}}{\Pr\big\{ y\,|\, x=1\big\}}.
\]
Since channels transmit symbols independently,
\[
\Pr\big\{ y\,|\, x=\alpha\big\}=\prod_{i=1}^{n}\Pr\big\{ y_{i}\,|\, x=\alpha\big\},\quad\alpha=0,1.
\]
Inserting the last formula in expression for $l(x)$, we have
\[
l(x)=\prod_{i=1}^{n}\frac{\Pr\big\{ y\,|\, x=0\big\}}{\Pr\big\{ y\,|\, x=1\big\}}=\prod_{i=1}^{n}l_{i}(x),
\]
or taking logarithms,
\[
L(x)=\sum_{i=1}^{n}L_{i}(x).
\]
Obtained formula gives the estimate of a bit
for which we have several independent sources of information.

\subsection{Estimation of a bit entering a linear system}

Let random vector variable $x=[x_{1},x_{2},\ldots,x_{n}]^{T}$
whose components take values $0$ and $1$ equiprobably, satisfy the linear system
\[
Ax=0,
\]
where the matrix $A\in\GF(2)^{m\times n}$ is exactly known.
Assume that the quantities $x_{1},\ldots,x_{n}$ are transmitted via
channels $W_{1},\ldots,W_{n}$, and received symbols are
$y=[y_{1},\ldots,y_{n}]$.
Then initial LLRs are known
\[
\lambda_{i}=\ln\frac{W_{i}(y_{i}|0)}{W_{i}(y_{i}|1)},\quad i=\overline{1,n}.
\]
Assume the following LLR is required
\[
L(x_{1})=\ln\frac{\Pr\{y|x_{1}=0\}}{\Pr\{y|x_{1}=1\}}
\]
without knowledge of $x$.

Note that if some component $x_i$ is exactly known, we can assume that 
it is transmitted via binary symmetric channel with zero error probability, and
\[
\lambda_{i}=\begin{cases}
+\infty, & x_{i}=0,\\
-\infty, & x_{i}=1.
\end{cases}
\]
Vice versa, if some bit $x_i$ is not transmitted, we can assume that
it is transmitted via absolutely noisy channel with the transition function
\[
W(0|a)=1,\quad a=0,1.
\]
It is easy to see that in this case
\[
\lambda_{i}=0.
\]
If such bit enters only one equation, we can remove this bit and respective equation.
If some equation contains only exactly known bits (with$\lambda_{i}=\pm\infty$),
this equation also can be removed.

Associate matrix $A$ with a bipartite undirected graph by the following rule.
Each matrix row (i.e. each equation) is associated with a square vertex.
Each matrix column (i.e. each bit variable) is associated with a round vertex.
A round vertex and a square vertex are connected by an edge if respective
matrix row and matrix column intersect at value one (i.e. if the respective
variable enters respective equation). Such a graph is referred to as Tanner graph for
the matrix $A$.

We focus on just one bit variable, say $x_1$.
If the Tanner graph is disconnected, remove all connected components
save one containing the vertex $x_1$.
Now if removing some bit $x_p$ results in emerging of $q$ graph components
 $A_{1},A_{2},\ldots,A_{q}$, our problem of estimating $x_1$ is split into
$q$ smaller problems. Let  $Y_{i}$
be a subvector of $y$ containing only those components which arise in transmission
of bits entering the subgraph $A_i$. Assume $p\neq1$, and let $x_{1}\in A_{1}$.
We can assume that $x_p$ is transmitted via $q-1$ different channels with transition functions
\[
\hat{W}_{i}(Y_{i}\,|\, a)=\Pr\Big\{\, Y_{i}\big|\, x_{p}=a\Big\},\quad i=\overline{2,q}.
\]
Compute LLRs of $x_p$ considering only channel $i$,
\[
L_{i}(x_{p})=\ln\frac{\hat{W}_{i}(Y_{i}\,|\,0)}{\hat{W}_{i}(Y_{i}\,|\,1)}.
\]
Then the subgraphs $A_{2},A_{3},\ldots,A_{q}$ may be removed
with updating the initial $\lambda_{p}$ estimate to
\[
\hat{\lambda}_{p}=\lambda_{p}+\sum_{i=1}^{q}L_{i}(x_{p}).
\]
Note that for each $i$ the problem of computation of  $L_{i}(x_{p})$ is also
a bit estimation problem formulated on a smaller graph  $A_{i}$ augmented by the vertex $x_p$.

\begin{figure}[b]
\begin{centering}
\includegraphics{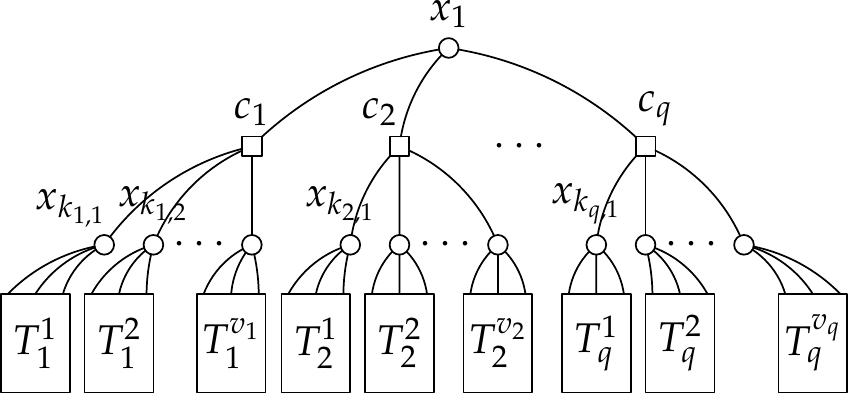}
\par\end{centering}

\caption{Tree-like Tanner graph given in the rooted tree form\label{fig:tanner-tree}}
\end{figure}

Now assume that the Tanner graph is a tree, i.e. it is connected and acyclic.
Assume also that every equation contains at least two variables.
Choose the vertex $x_1$ as a tree root vertex. The leaf vertices will be some subset
of $x_{2},\ldots,x_{n}$. Let the vertex $x_1$ be incident to equations  $c_{1},c_{2},\ldots,c_{q}$,
and let each vertex  $c_{i},\; i=\overline{1,q}$ be incident to variables
$x_{k_{i,1}},\ldots x_{k_{i,v_{i}}}$, not counting $x_1$.
Let $T_{i}^{j}$ be the subtree with root at $x_{k_{i,j}}$, not counting the root itself
(see fig.~\ref{fig:tanner-tree}).

Let $I_{i}^{j}\subset\{1,2,\ldots,n\}$ be an index set for variables
entering the subgraph $T_{i}^{j}$. For all $i,j$ define the set
\[
Y_{i}^{j}=\{y_{r}:\: r\in I_{i}^{j}\}\cup\{y_{k_{i,j}}\},
\]
i.e. the set of all symbols obtained via transmitting variables entering
the subtree rooted at $x_{k_{i,j}}$. Assume that for each pair $i,j$ we know the LLRs
\begin{equation}
L(x_{k_{i,j}})=\ln\frac{\Pr\big\{ Y_{i}^{j}\,|\, x_{k_{i,j}}=0\big\}}{\Pr\big\{ Y_{i}^{j}\,|\, x_{k_{i,j}}=1\big\}},\label{eq:lxkij}
\end{equation}
i.e. bit $x_{k_{i,j}}$ estimates based only on tree rooted at the vertex $x_{k_{i,j}}$.
This can be interpreted as transmitting each such bit via the channel with the following
transition function,
\[
W'_{ij}(Y_{i}^{j}\,|\, a)=\Pr\big\{ Y_{i}^{j}\,|\, x_{k_{i,j}}=a\big\}.
\]
Write each equation $c_i$ in the following way:
\begin{equation}
x_{1}=x_{k_{i,1}}\oplus x_{k_{i,2}}\oplus\ldots\oplus x_{k_{i,v_{i}}},\quad i=\overline{1,q}.\label{eq:x1eq}
\end{equation}
We can assume that $x_1$ was transmitted via independent channels $\hat{W}_{i}$
with transition functions
\[
\hat{W}_{i}(Y_{i}^{1},Y_{i}^{2},\ldots,Y_{i}^{l_{i}}|a)=\Pr\big\{ Y_{i}^{1},Y_{i}^{2},\ldots,Y_{i}^{l_{i}}|x_{1}=a\big\},\quad i=\overline{1,q}.
\]
Then LLRs based on these channels have the form
\[
L_{i}(x_{1})=\ln\frac{\Pr\big\{ Y_{i}^{1},Y_{i}^{2},\ldots,Y_{i}^{l_{i}}|x_{1}=0\big\}}{\Pr\big\{ Y_{i}^{1},Y_{i}^{2},\ldots,Y_{i}^{l_{i}}|x_{1}=1\big\}}.
\]
Inserting (\ref{eq:x1eq}), we get
\[
L_{i}(x_{1})=\ln\frac{\Pr\big\{ Y_{i}^{1},Y_{i}^{2},\ldots,Y_{i}^{l_{i}}|x_{k_{i,1}}\oplus x_{k_{i,2}}\oplus\ldots\oplus x_{k_{i,v_{i}}}=0\big\}}{\Pr\big\{ Y_{i}^{1},Y_{i}^{2},\ldots,Y_{i}^{l_{i}}|x_{k_{i,1}}\oplus x_{k_{i,2}}\oplus\ldots\oplus x_{k_{i,v_{i}}}=1\big\}}.
\]
Taking into account (\ref{eq:lxkij}) and using result of section \ref{sub:check-node-update},
we obtain
\begin{eqnarray*}
L_{i}(x_{1})&=&L(x_{k_{i,1}}\oplus x_{k_{i,2}}\oplus\ldots\oplus x_{k_{i,v_{i}}})\\
&=&L(x_{k_{i,1}})\llrbox L(x_{k_{i,2}})\llrbox\ldots\llrbox L(x_{k_{i,v_{i}}}).
\end{eqnarray*}
Finally assuming $x_1$ be transmitted via the channels $\hat{W}_{1},\hat{W}_{2},\ldots,\hat{W}_{q}$
and also via $W_1$, write
\[
L(x_{1})=\lambda_{1}+\sum_{i=1}^{q}L_{i}(x_{1}).
\]
In order to compute $L(x_{k_{i,j}})$, we can apply the same reasoning to the subtree
rooted at $x_{k_{i,j}}$. Thus we have a recursive algorithm computing  $L(x_{1})$.
It is essentially equivalent to the algorithm known as ``belief propagation''.

\section{Polar codes\label{sec:polar}}

\begin{figure}[h]
\begin{centering}
\includegraphics{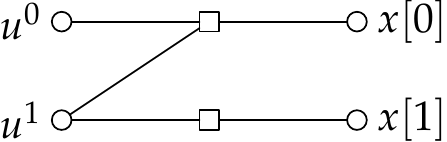}
\par\end{centering}

\caption{The encoder graph\label{fig:coder-graph}}
\end{figure}

In this section we consider polar codes in exactly that form which they were
presented in originally \cite{DBLP:journals/corr/abs-0807-3917},
but take a slightly different look.
Let $u^0$ and $u^1$ be two independent random bits taking values $0$ and $1$ equiprobably.
Define two more bits
\begin{eqnarray}
x[0] & = & u^{0}\oplus u^{1},\nonumber \\
x[1] & = & u^{1}.\label{eq:x0x1}
\end{eqnarray}
In matrix notation,
\[
[x[0],x[1]]=[u^{0},u^{1}]\cdot G_{2},\quad\mbox{где }G_{2}=\left[\begin{array}{cc}
1 & 0\\
1 & 1
\end{array}\right].
\]
Note that bits  $x[0]$, $x[1]$ also take the values $0$ and $1$ equiprobably.
Construct the Tanner graph for the system (\ref{eq:x0x1})
and denote it as \emph{encoder graph} (see fig.~\ref{fig:coder-graph}).

\begin{figure}[h]
\begin{centering}
\includegraphics{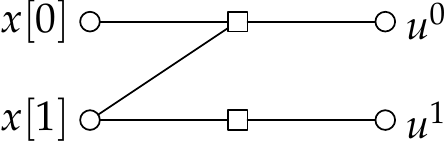}
\par\end{centering}

\caption{The decoder graph\label{fig:decoder-graph}}
\end{figure}

Since $G_{2}^{-1}=G_{2}$, we can rewrite the system (\ref{eq:x0x1})
in equivalent form
\begin{eqnarray*}
u^{0} & = & x[0]\oplus x[1],\\
u^{1} & = & x[1].
\end{eqnarray*}
Construct the Tanner graph for this system also and denote it as
\emph{decoder graph} (see fig.~\ref{fig:decoder-graph}).

Let bits $x[0]$ and $x[1]$ be transmitted via a given channel $W$
receiving symbols  $y=[y^{0},y^{1}]$.
Assume the following LLR is required
\[
L(u^{0})=\ln\frac{\Pr\big\{ y\,|\, u^{0}=0\big\}}{\Pr\big\{ y\,|\, u^{0}=1\big\}}.
\]
Using results of the section~\ref{sec:bp-on-trees}, we have
\[
L(u^{0})=L(x[0])\llrbox L(x[1]),
\]
where
\[
L(x[i])=\ln\frac{W(y^{i}\,|\,0)}{W(y^{i}\,|\,1)},\quad i=0,1.
\]
Now assume that the value of $u^{0}$ is exactly known and that we need
\[
L(u^{1})=\ln\frac{\Pr\big\{ y\,|\, u^{1}=0,\, u^{0}\big\}}{\Pr\big\{ y\,|\, u^{1}=1,\, u^{0}\big\}}.
\]
Using again section~\ref{sec:bp-on-trees}, we get
\[
L(u^{1})=L(x[1])+(L(x[0])\llrbox L(u^{0}))=L(x[1])+(-1)^{u^{0}}L(x[0]).
\]
We proceed to recursive construction of the larger system and then to similar problem
of determining of one bit.

\subsection{Hierarchical graph construction}

\begin{figure}[h]
\begin{centering}
\includegraphics{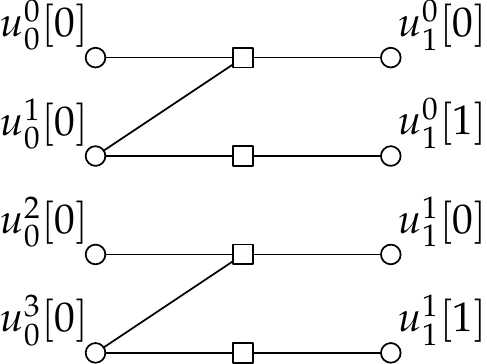}
\par\end{centering}

\caption{Graph for the system of equations (\ref{eq:duplicated})\label{fig:duplicated}}
\end{figure}

Double the encoder graph taking two copies of each variable and of each equation.
Now let $u^{0},u^{1},u^{2},u^{3}$ be the transmitted random bits while
$u_{1}^{0}[0],u_{1}^{0}[1],u_{1}^{1}[0],u_{1}^{1}[1]$ are their functions:
\begin{eqnarray}
u_{1}^{0}[0] & = & u^{0}\oplus u^{1},\nonumber \\
u_{1}^{0}[1] & = & u^{1},\nonumber \\
u_{1}^{1}[0] & = & u^{2}\oplus u^{3},\nonumber \\
u_{1}^{1}[1] & = & u^{3}.\label{eq:duplicated}
\end{eqnarray}
From the other hand,
\begin{eqnarray*}
u^{0} & = & u_{1}^{0}[0]\oplus u_{1}^{0}[1],\\
u^{1} & = & u_{1}^{0}[1],\\
u^{2} & = & u_{1}^{1}[0]\oplus u_{1}^{1}[1],\\
u^{3} & = & u_{1}^{1}[1].
\end{eqnarray*}
For consistency, set $u_{0}^{i}[0]\equiv u^{i}$. Figure~\ref{fig:duplicated}
gives the graph for the system  (\ref{eq:duplicated}).

\begin{figure}[h]
\begin{centering}
\includegraphics{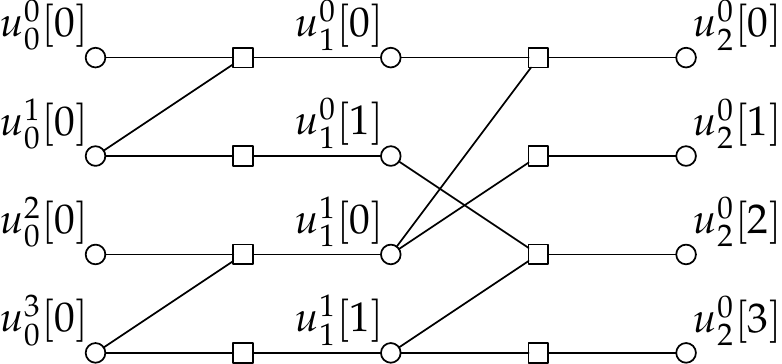}
\par\end{centering}

\caption{Graph for the system of equations (\ref{eq:coder-2})\label{fig:coder-2}}
\end{figure}

Introduce four new variables (fig.~\ref{fig:coder-2}),
\begin{eqnarray}
u_{2}^{0}[0] & = & u_{1}^{0}[0]\oplus u_{1}^{1}[0],\nonumber \\
u_{2}^{0}[1] & = & u_{1}^{1}[0],\nonumber \\
u_{2}^{0}[2] & = & u_{1}^{0}[1]\oplus u_{1}^{1}[1],\nonumber \\
u_{2}^{0}[3] & = & u_{1}^{1}[1].\label{eq:coder-2}
\end{eqnarray}
In matrix notation,
\begin{eqnarray*}
[u_{2}^{0}[0],u_{2}^{0}[1]] & = & [u_{1}^{0}[0],u_{1}^{1}[0]]\cdot G_{2},\\
{}[u_{2}^{0}[2],u_{2}^{0}[3]] & = & [u_{1}^{0}[1],u_{1}^{1}[1]]\cdot G_{2}.
\end{eqnarray*}

\begin{figure}[h]
\begin{centering}
\includegraphics{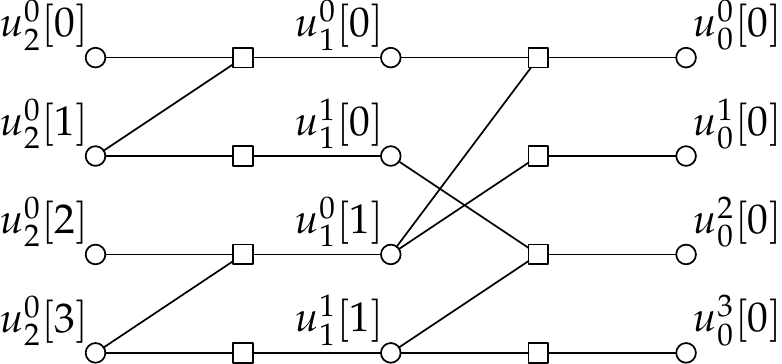}
\par\end{centering}

\caption{Graph for the system of equations (\ref{eq:decoder-2})\label{fig:decoder-2}}
\end{figure}

Again employ the relation $G_{2}^{-1}=G_{2}$ and express old variables in terms of new ones
(fig.~\ref{fig:decoder-2}),
\begin{eqnarray}
u_{1}^{0}[0] & = & u_{2}^{0}[0]\oplus u_{2}^{0}[1],\nonumber \\
u_{1}^{1}[0] & = & u_{2}^{0}[1],\nonumber \\
u_{1}^{0}[1] & = & u_{2}^{0}[2]\oplus u_{2}^{0}[3],\nonumber \\
u_{1}^{1}[1] & = & u_{2}^{0}[3],\label{eq:decoder-2}
\end{eqnarray}

We call the graph displayed on fig.~\ref{fig:coder-2} the encoder graph
and the graph displayed on fig.~\ref{fig:decoder-2} the decoder graph.
Now we are able to repeat the whole operation, double the graph and introduce
eight new variables  $u_{3}^{0}[i],\, i=\overline{0,7}$, and proceed further.
We can express new variables in terms of old ones,
\begin{equation}
[u_{k+1}^{i}[2j],u_{k+1}^{i}[2j+1]]=[u_{k}^{2i}[j],u_{k}^{2i+1}[j]]\cdot G_{2},\label{eq:new-old}
\end{equation}
and vice versa:
\[
[u_{k}^{2i}[j],u_{k}^{2i+1}[j]]=[u_{k+1}^{i}[2j],u_{k+1}^{i}[2j+1]]\cdot G_{2}.
\]
Assume we make $n$ steps and stop at introducing new variables
$u_{n}^{0}[i]$. Indices in the expression  $u_{k}^{i}[j]$ have the following
interpretation based on decoder graph.
Lower index $k$ specifies the vertical ``layer'' of the graph of $2^n$ variables
where the given vertex is, if we count layers right to left.
Bracketed index $j$ specifies the independent group of variables in a layer.
Upper index $i$ specifies the variable inside a group.

\begin{figure}
\begin{centering}
\includegraphics[scale=0.7]{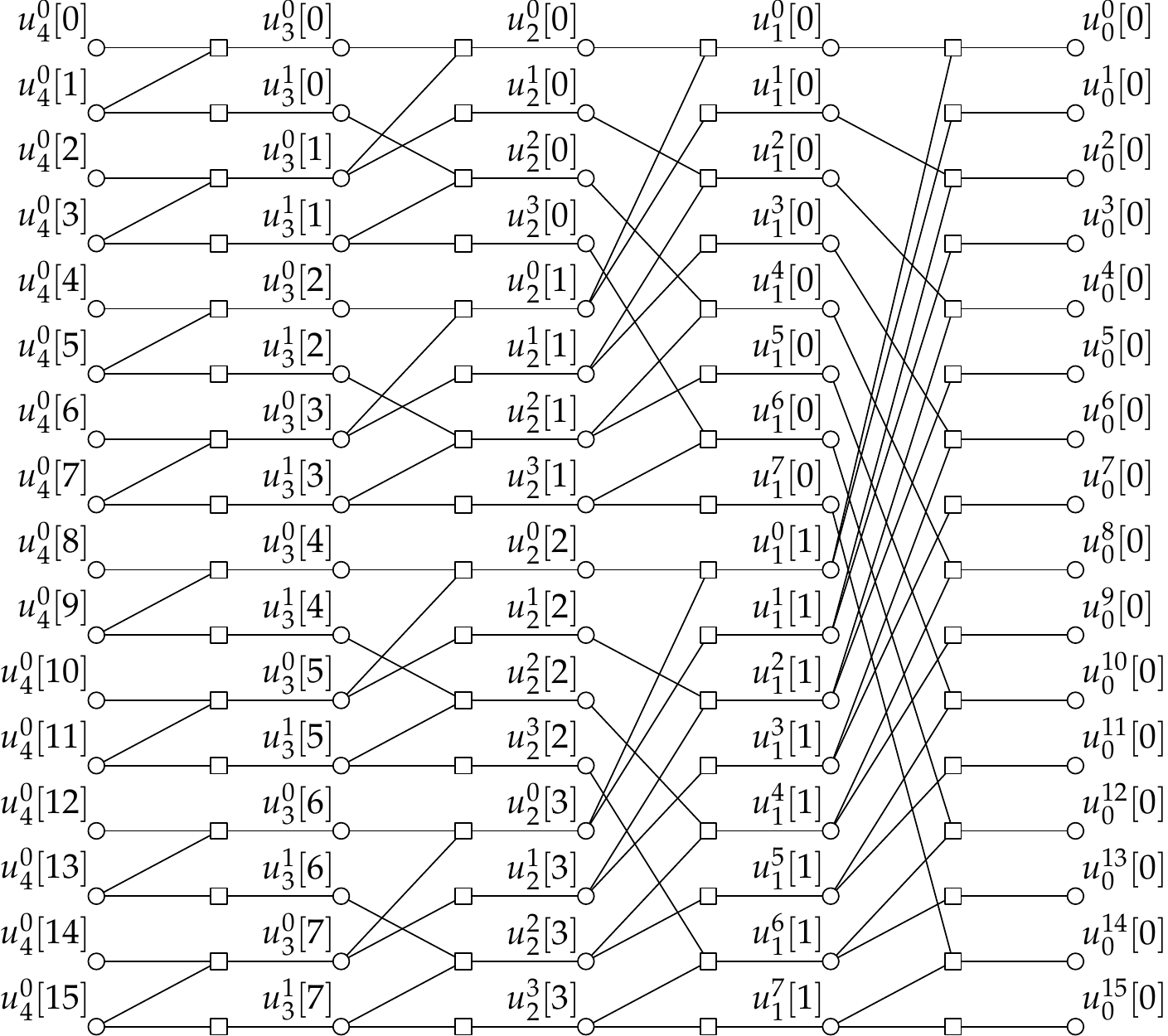}
\par\end{centering}
\caption{Decoder graph for $n=4$.\label{fig:decoder-4}}
\end{figure}

Increasing layer number by one doubles the number of independent groups
$J_{k}$, i.e.
\[
J_{k}=2J_{k-1},\quad k=\overline{1,n}.
\]
Layer indexed $0$ contains one group of $2^n$ variables, i.e.
$J_{0}=1$. Hence
\[
J_{k}=2^{k}.
\]
Since every layer contains $2^n$ variables, the number of variables in every
group of layer $k$ is
\[
S_{k}=\frac{2^{n}}{J_{k}}=2^{n-k}.
\]
Thus, upper index in the expression $u_{k}^{i}[j]$ has range $0$ to  $S_{k}-1$,
while bracketed index has range  $0$ to $J_{k}-1$.
Figure~\ref{fig:decoder-4} shows decoder graph for $n=4$.

\subsection{Problem of bit estimation}

Let the variables $u_{n}^{0}[j]$ constituting the graph last layer,
be transmitted via some channel $W$ and received as symbols
$y=[y_{0},y_{1},\ldots,y_{J_{n}-1}].$ Using the channel model, we have
\[
\lambda_{j}=\ln\frac{W(y_{j}\,|\,0)}{W(y_{j}\,|\,1)}.
\]
Assume that for some $m$ we exactly know the quantities
\[
u_{0}^{0}[0],u_{0}^{1}[0],u_{0}^{2}[0],\ldots,u_{0}^{m-1}[0].
\]
Assume that the estimate of the next bit  $u_{0}^{m}[0]$ is required, i.e.
\begin{equation}
L(u_{0}^{m}[0])\equiv\ln\frac{\Pr\big\{ y\,|\, u_{0}^{m}[0]=0,\, u_{0}^{i}[0],i=\overline{0,m-1}\big\}}{\Pr\big\{ y\,|\, u_{0}^{m}[0]=1,\, u_{0}^{i}[0],i=\overline{0,m-1}\big\}}.\label{eq:final-llr}
\end{equation}
Denote the subvector of $y$ consisting of contiguous bits
from index $j\cdot S_{k}$ up to index $(j+1)\cdot S_{k}-1$ inclusively
by symbol  $Y_{k}[j]$.
Then the vector $y$ has the following representation in terms of
its parts  $Y_{k}[j]$,
\[
y=\big[Y_{k}[0],Y_{k}[1],\ldots,Y_{k}[J_{k}]\big].
\]
On the decoder graph, the subvector $Y_{k}[j]$ may be interpreted
as components of $y$ corresponding to those bits $u_{n}^{0}[l]$ which
are strictly on the left of variables of group $j$, layer $k$.
For example, if $n=4$ (fig.~\ref{fig:decoder-4}) then group $1$ of layer $2$
consists of variables $u_{2}^{0}[1],u_{2}^{1}[1],u_{2}^{2}[1],u_{2}^{3}[1],$
while the subvector $Y_{2}[1]$ is obtained via transmission of the bits
 $u_{4}^{0}[4],u_{4}^{0}[5],u_{4}^{0}[6],u_{4}^{0}[7]$.
Introduce the following notation,
\begin{equation}
L(u_{k}^{i}[j])\equiv\ln\frac{\Pr\big\{ Y_{k}[j]\,|\, u_{k}^{i}[j]=0,\, u_{k}^{l}[j],l=\overline{0,i-1}\big\}}{\Pr\big\{ Y_{k}[j]\,|\, u_{k}^{i}[j]=1,\, u_{k}^{l}[j],l=\overline{0,i-1}\big\}}.\label{eq:intermediate-llr}
\end{equation}
This means that $L(u_{k}^{i}[j])$ is the LLR for the bit $u_{k}^{i}[j]$ provided
that  $Y_{k}[j]$ is received and that the quantities
\[
u_{k}^{0}[j],u_{k}^{1}[j],\ldots,u_{k}^{i-1}[j]
\]
are exactly known. Note that the formula (\ref{eq:final-llr}) is
a special case of (\ref{eq:intermediate-llr}) for $k=0$,
and that for $k=n$ the formula (\ref{eq:intermediate-llr}) takes the form
\begin{equation}
L(u_{n}^{0}[j])\equiv\ln\frac{\Pr\big\{ y_{j}\,|\, u_{n}^{0}[j]=0\big\}}{\Pr\big\{ y_{j}\,|\, u_{n}^{0}[j]=1\big\}}=\lambda_{j}.\label{eq:llr-base}
\end{equation}

Our goal is to obtain recursive formula for $L(u_{k}^{i}[j])$
in terms of  $L(u_{k+1}^{m}[l]).$
If we have it, we can compute the required $L(u_{0}^{m}[0])$ using
(\ref{eq:llr-base}) as a recursion base.

Denote the subgraph consisting of single vertex $u_{n}^{0}[j]$
by  $A_{n}[j]$.
By induction, let  $A_{k}[j]$ for $j=\overline{0,J_{k}-1}$
be the union of subgraphs  $A_{k+1}[2j],A_{k+1}[2j+1],$
of all vertices of group $j$, layer $k$ and of incident equations.
On the graph drawing we can interpret $A_{k+1}[j]$ as a subgraph
whose vertices are all bits of group $j$, layer $k$ and all vertices
on the left of these. For example, if $n=4$, the subgraph
$A_{2}[1]$ consists of the variables
\[
\begin{array}{cccc}
u_{4}^{0}[4], & u_{4}^{0}[5], & u_{4}^{0}[6], & u_{4}^{0}[7],\\
u_{3}^{0}[2], & u_{3}^{1}[2], & u_{3}^{0}[3], & u_{3}^{1}[3],\\
u_{2}^{0}[1], & u_{2}^{1}[1], & u_{2}^{2}[1], & u_{2}^{3}[1]
\end{array}
\]
and incident equations (fig.~\ref{fig:decoder-4}).
Note that the subgraph $A_{k}[j]$ contains those and only those variables
of layer $n$, whose transmission results in the vector $Y_{k}[j]$.

\subsection{Recursive formulas}

\begin{figure}
\begin{centering}
\includegraphics[scale=0.7]{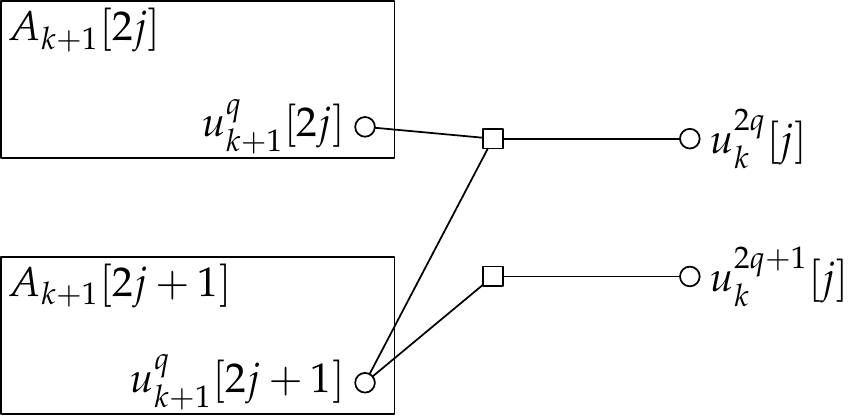}
\par\end{centering}
\caption{The decoder graph after vertex removal\label{fig:two-subgraphs}}
\end{figure}

Here we find the expression for $L(u_{k}^{i}[j])$ under the constraint $k<n$.
All variables of layer $0$ enter only one equation, and for $k>0$
we do not have any immediate information for them, thus we remove them and the incident equation.
Now for $k>1$ the same can be done for layer $1$ etc. Finally we retain only layers with index
at least $k$.
The graph will be divided in $J_k$ connected components $A_{k}[l]$.
Remove all components save one containing  $u_{k}^{i}[j]$,
which means that we keep only the component $A_{k}[j]$. Denote $q=\lfloor i/2\rfloor$.
If bits  $u_{k}^{0}[j],u_{k}^{1}[j],\ldots,u_{k}^{i-1}[j]$
are exactly known, then according to the equation (\ref{eq:new-old}),
exactly known are also the quantities
\begin{equation}
u_{k+1}^{l}[2j],\, u_{k+1}^{l}[2j+1],\quad l=\overline{0,q-1}.\label{eq:u-k-plus-1}
\end{equation}
Hence the equations incident to $u_{k}^{0}[j],u_{k}^{1}[j],\ldots,u_{k}^{i-1}[j]$
and to (\ref{eq:u-k-plus-1}) may be removed: these equations contain only known quantities.
After the removal the vertices
\[
u_{k}^{0}[j],u_{k}^{1}[j],\ldots,u_{k}^{2q-1}[j]
\]
become isolated and also may be removed. The vertices
\[
u_{k}^{2q+2}[j],u_{k}^{2q+3}[j],\ldots,u_{k}^{S_{k}-1}[j]
\]
are not transmitted and do not have any estimates, thus also may be removed
with corresponding equations (one per vertex).
After these transformations the graph will have the form depicted on fig.~\ref{fig:two-subgraphs}.

\begin{figure}
\begin{centering}
\includegraphics{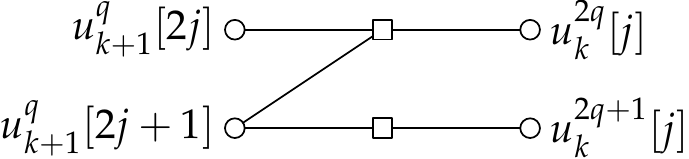}
\par\end{centering}

\caption{The graph after substitution of vertices
$u_{k+1}[2j]$, $u_{k+1}[2j+1]$ instead of subgraphs
$A_{k+1}[2j]$, $A_{k+1}[2j+1]$, respectively.\label{fig:simplified}}
\end{figure}

Since removing the vertex $u_{k+1}^{q}[2j]$ divides the graph into two components
one of which is $A_{k+1}[2j]$, we can assume that the bit $u_{k+1}^{q}[2j]$
is transmitted via the channel with the following transition function
\begin{equation}
\begin{array}{rcl}
\hat{W}_{0}(Y_{k+1}[2j]\,|\, a)&=&\Pr\Big\{ Y_{k+1}[2j]\,\big|\, u_{k+1}^{q}[2j]=a;\\
&& u_{k+1}^{l}[2j+i],l=\overline{0,q-1}\Big\}.
\end{array}\label{eq:w0-k-plus-1}
\end{equation}
Also we can substitute the subgraph $A_{k+1}[2j]$ by the single vertex  $u_{k+1}^{q}[2j]$,
with the following initial likelihood ratio,
\[
\lambda_{k+1}^{q}[2j]=\ln\frac{\hat{W}_{0}(Y_{k+1}[2j]\,|\,0)}{\hat{W}_{0}(Y_{k+1}[2j]\,|\,1)}.
\]
Comparing (\ref{eq:intermediate-llr}) with (\ref{eq:w0-k-plus-1}),
we conclude that
\[
\lambda_{k+1}^{q}[2j]=L(u_{k+1}^{q}[2j]).
\]
Similarly, the vertex $u_{k+1}^{q}[2j+1]$ can be substituted for the whole subgraph
$A_{k+1}[2j+1]$, if we set
\[
\lambda_{k+1}^{q}[2j+1]=L(u_{k+1}^{q}[2j+1]).
\]
Resulting graph is displayed on fig.~\ref{fig:simplified}.
We know already that for such graph,
\begin{equation}
\begin{array}{rcl}
L(u_{k}^{2q}[j]) & = & L(u_{k+1}^{q}[2j])\llrbox L(u_{k+1}^{q}[2j+1]), \\
L(u_{k}^{2q+1}[j]) & = & L(u_{k+1}^{q}[2j+1])+(-1)^{u_{k}^{2q}[j]}L(u_{k+1}^{q}[2j]).
\end{array}
\label{eq:llr-recursion}
\end{equation}

Thus we have obtained the recursive formulas giving
$L(u_{0}^{i}[0])$ for all $i=\overline{0,2^{n}-1}$ using (\ref{eq:llr-base})
as a recursion base.

\subsection{Successive cancellation method}

Let the vector
\[
u=\big[u_{0}^{0}[0],u_{0}^{1}[0],u_{0}^{2}[0],\ldots,u_{0}^{2^{n}-1}[0]\big]
\]
be the message required for transmission.
Starting from $u$ we can compute
\[
x=\big[u_{n}^{0}[0],u_{n}^{0}[1],u_{n}^{0}[2],\ldots,u_{n}^{0}[2^{n}-1]\big],
\]
using the formulas (\ref{eq:new-old}). The vector $x$ will be considered as a codeword
and transmitted componentwise via given symmetric channel $W$ producing vector $y$
at the receiver. We want to recover $u$ using $y$. We do it sequentially bit by bit.
First we compute $L(u_{0}^{0}[0])$ and estimate the bit $u_{0}^{0}[0]$
as follows,
\[
\hat{u}_{0}^{0}[0]=\begin{cases}
0, & L(u_{0}^{0}[0])>0,\\
1, & L(u_{0}^{0}[0])<0,\\
\mbox{choose randomly,} & L(u_{0}^{0}[0])=0.
\end{cases}
\]
Now assuming  $u_{0}^{0}[0]$ exactly known, we compute
$L(u_{0}^{1}[0])$, estimate $u_{0}^{1}[0]$ etc. Each bit
is estimated using the rule
\begin{equation}
\hat{u}_{0}^{i}[0]=\begin{cases}
0, & L(u_{0}^{i}[0])>0,\\
1, & L(u_{0}^{i}[0])<0,\\
\mbox{choose randomly,} & L(u_{0}^{i}[0])=0.
\end{cases}\label{eq:hard-decision}
\end{equation}
Finally we produce some estimate of the initial message $u$.
This decoding method is called \emph{successive cancellation}.
Of course, presented coding system is useless since the redundancy is missing.

\subsection{Polar codes}

Choose some index set
$F\subset\{0,1,2,\ldots,2^{n}-1\}.$
Denote $K=2^{n}-|F|.$ Make a convention that the only possible messages are those
with bits from $F$ equal to zero. We call these bits \emph{frozen}, and other ones
\emph{information} bits.
Again we use the successive cancellation however with modified bit estimation rule:
\[
\hat{u}_{0}^{i}[0]=\begin{cases}
0, & i\in F,\\
\mbox{choose using (\ref{eq:hard-decision}),} & \mbox{otherwise.}
\end{cases}
\]
Since the admissible messages form the linear space of dimension $K$
and codewords $x$ depend linearly on $u$, the set of all codewords
is also a linear space. In other words, we have linear block code of length
$2^n$ and of rate $K/2^n$.
It can be shown that its generator is obtained by deleting rows with indices in $F$
from the matrix
\[
G_{2}^{\otimes n}\cdot R_{n},
\]
where $G_{2}^{\otimes n}$ denotes the $n$-th Kronecker power of the matrix $G_2$
and $R_n$ the \emph{bit reverse permutation} matrix.
The rule describing this permutation is as follows: let binary representation of index $i$
be $\alpha_{n-1}\alpha_{n-2}\ldots\alpha_{2}\alpha_{1}\alpha_{0}$, then 
the element $i$ is swapped with the element indexed
$\alpha_{0}\alpha_{1}\alpha_{2}\ldots\alpha_{n-2}\alpha_{n-1}$, in binary representation.
Thus constructed code is called the \emph{polar code.}

How to choose the set $F$ of frozen bits?
Denote the probability of erroneous detection of the bit $u_{0}^{i}[0]$
using the successive cancellation method provided that all previous bits are detected
correctly and $F=\varnothing,$ by $E_{i},i=\overline{0,2^{n}-1}$.
The probability of block error $P_E$ with $F$ fixed may be estimated from above
as a sum of probability errors for each information bit, i.e.
\begin{equation}
P_{E}\leq\sum_{i\notin F}E_{i}.\label{eq:fer-upper-bound}
\end{equation}
The set $F$ may contain indices of bits with maximal error probabilities,
which will minimize the upper bound (\ref{eq:fer-upper-bound}) of the block error probability.
To this end, one has to compute the probabilities $E_{i}$, which is discussed in section~\ref{sec:design}.

\subsection{Complexity of encoding and decoding}

Polar codes would not have any practical value without fast algorithms of encoding and decoding.
The encoding process is carried out by recursive formulas
(\ref{eq:new-old})
\[
[u_{k+1}^{i}[2j],u_{k+1}^{i}[2j+1]]=[u_{k}^{2i}[j],u_{k}^{2i+1}[j]]\cdot G_{2}
\]
and requires $n$ sequential steps $k=1,2,3,\ldots,n$. On each of these steps,
all variables of layer $k$ are defined. Since each layer contains $2^n$ bits, the overall
encoding complexity is $O(n\cdot2^n)$ operations, which is $O(N\cdot\log_2N)$ if
we introduce the code length $N=2^n$.

Decoding by successive cancellation method using the recursive formulas
(\ref{eq:llr-recursion}) requires computation of  $n\cdot2^{n}$
different quantities $L(u_{k}^{i}[j])$ and of $n\cdot2^{n}$
quantities $u_{k}^{i}[j]$ in a more complex order. Hence the decoding complexity
also is  $O(N\cdot log_{2}N)$ operations.

\section{Construction and analysis of polar codes\label{sec:design}}

Construction of polar code of given length $N=2^n$ and rate  $\frac{K}{2^{n}}$
for a given channel $W$ amounts to choosing the set $F$ of $N-K$ frozen bits.
The choice which minimizes the block error probability $P_E$ would be optimal.
However computation of $P_E$ is complicated  and it is reasonable to
substitute its upper bound in the minimization problem,
\[
\min_F\sum_{i\notin F}E_{i},
\]
where $E_i$ is the probability of erroneous detection of bit $i$ by successive cancellation
under assumption that all previous bits are detected without error.
In this formulation, it is sufficient to choose $N-K$ indices corresponding to maximal
values of $E_i$ as the set $F$ provided that $E_i$ are known for $i=\overline{0,2^{n}-1}$.
Thus the polar code construction problem is reduced to computation of quantities $E_i$.

\subsection{Likelihood ratios as random variables}

Since the channel is symmetric and the code is linear, in computation of $E_i$
we can assume that all-zeros codeword is transmitted.
In this case the probability to receive the vector $y$ is
\begin{equation}
p(y)=\prod_{i=0}^{N-1}W(y_{i}\,|\,0).\label{eq:prob-measure}
\end{equation}
By definition of $E_i$ we assume all bits $u_{0},u_{1},\ldots,u_{i-1}$ zero. In this case
the recursion (\ref{eq:llr-recursion}) takes the form
\begin{eqnarray}
L(u_{k}^{2q}[j]) & = & L(u_{k+1}^{q}[2j])\llrbox L(u_{k+1}^{q}[2j+1]),\nonumber \\
L(u_{k}^{2q+1}[j]) & = & L(u_{k+1}^{q}[2j+1])+L(u_{k+1}^{q}[2j]).\label{eq:llr-recursion-zero}
\end{eqnarray}
The recursion base (\ref{eq:llr-base}) remains unchanged:
\[
L(u_{n}^{0}[j])=\lambda_{j}=\ln\frac{W(y_{j}\,|\,0)}{W(y_{j}\,|\,1)}.
\]
Now the quantities $L(u_{k}^{i}[j])$ depend only on $y$ and do not depend on
$u=[u_{0}^{0}[0],\ldots,u_{0}^{N-1}[0]]$, thus we consider $L(u_{k}^{i}[j])$ as random
variables defined on probability space $\mathcal{Y}^{N}$ with probability measure (\ref{eq:prob-measure}).

According to (\ref{eq:prob-measure}), the quantities  $y_{0},y_{1},\ldots,y_{N-1}$ are mutually
independent. Hence the quantities  $L(u_{n}^{0}[0]),L(u_{n}^{0}[1]),\ldots,L(u_{n}^{0}[N-1])$
are also mutually independent, because every quantity  $L(u_{n}^{0}[j])$ depends on only one
symbol $y_j$. The quantities $L(u_{k}^{i'}[j'])$
and  $L(u_{k}^{i''}[j''])$ are independent for all $k>0$, $i'$, $i''$ and $j'\ne j''$,
because they are defined by recursive formulas (\ref{eq:llr-recursion-zero}) via
non-intersecting sets of $L(u_{n}^{0}[j])$.

Following the hard decision rule (\ref{eq:hard-decision}) we see that the bit $i$
is detected erroneously in all cases when $L(u_{0}^{i}[0])<0$
and in half of cases when $L(u_{0}^{i}[0])=0$. In other words, $E_i$
is%
\footnote{If $L(u_{0}^{i}[0])$ has continuous distribution, the term
$\frac{1}{2}\Pr\{L(u_{0}^{i}[0])=0\}$ should be deleted. }
\[
E_{i}=\Pr\big\{ L(u_{0}^{i}[0])<0\big\}+\frac{1}{2}\Pr\big\{ L(u_{0}^{i}[0])=0\big\}.
\]
Extend the problem of computation of $E_i$ to computation of distributions
of random variables $L(u_{0}^{i}[0])$. Denote by $f_{k}^{i}[j]$
the probability function%
\footnote{In the continuous case, $f_{k}^{i}[j]$ will be the probability density function. }
of the random variable $L(u_{k}^{i}[j])$:
\[
f_{k}^{i}[j](z)=\Pr\big\{ L(u_{k}^{i}[j])=z\big\}.
\]
From the channel model we have
 $f_{n}^{0}[j],j=\overline{0,N-1}$:
\begin{eqnarray*}
f_{n}^{0}[j](z) & = & \Pr\Big\{\ln\frac{W(y_{j}|0)}{W(y_{j}|1)}=z\,\big|\, u_{n}^{0}[j]=0\Big\}=\\
 & = & \sum_{b:\ln\frac{W(b|0)}{W(b|1)}=z}W(b\,|\,0).
\end{eqnarray*}
We see that the distributions $f_{n}^{0}[j]$ of random variables  $L(u_{n}^{0}[j])$
are the same for all $j$. Formulas (\ref{eq:llr-recursion-zero}) imply that
for $k<n$ the distributions of $L(u_{k}^{i}[j])$ also do not depend on $j$, i.e.
$f_{k}^{i}[j']=f_{k}^{i}[j'']\;\forall i,k,j',j''$.
Therefore in what follows, we drop the square brackets in the notation  $f_{k}^{i}[j]$.

\subsection{Recurrent relations for the distributions}
Here we show that the distributions  $f_{k}^{i}$ satisfy the recurrent relations
analogous to the formulas  (\ref{eq:llr-recursion-zero}).
We start from random variables with odd indices:
\[
L(u_{k}^{2q+1}[j])=L(u_{k+1}^{q}[2j+1])+L(u_{k+1}^{q}[2j]).
\]
Since
$L(u_{k+1}^{q}[2j])$ and $L(u_{k+1}^{q}[2j+1])$ are i.i.d.,
\begin{eqnarray*}
f_{k}^{2q+1}(z)&=&\Pr\big\{ L(u_{k}^{2q+1}[j])=z\big\}\\
&=&\sum_{a,b\,\in\,\supp f_{k+1}^{q}:\; a+b=z,}f_{k+1}^{q}(a)f_{k+1}^{q}(b),
\end{eqnarray*}
where $\supp f_{k+1}^{q}=\{v:f_{k+1}^{q}(v)\neq0\}.$ Rewrite the sum so that
it will go over one index only:
\begin{equation}
f_{k}^{2q+1}(z)=\sum_{a\,\in\,\supp f_{k+1}^{q}}f_{k+1}^{q}(a)f_{k+1}^{q}(z-a).\label{eq:f-2q-plus-1-z}
\end{equation}
If $\supp f_{k+1}^{q}$ is a uniform mesh, the formula
(\ref{eq:f-2q-plus-1-z}) is nothing else but discrete convolution of a sequence with itself.
In the continuous case, the formula (\ref{eq:f-2q-plus-1-z}) will have the form
\[
f_{k}^{2q+1}(z)=\int\limits _{-\infty}^{+\infty}f_{k+1}^{q}(a)f_{k+1}^{q}(z-a)da,
\]
which is also convolution of the function $f_{k+1}^{q}$ with itself.
Hence it is quite natural to call the probability function $h$ defined by the formula
\[
h(z)=\sum_{a\,\in\,\supp f}f(a)g(z-a),
\]
the convolution of the functions $f$ and $g$ with the notation
\[
h=f\star g.
\]
Thus in new notation
\begin{equation}
f_{k}^{2q+1}=f_{k+1}^{q}\star f_{k+1}^{q}.\label{eq:f-odd}
\end{equation}
Random variables with even indices are treated analogously. Using the corresponding
recursive formula
\[
L(u_{k}^{2q}[j])=L(u_{k+1}^{q}[2j])\llrbox L(u_{k+1}^{q}[2j])
\]
we write
\begin{eqnarray*}
f_{k}^{2q}(z)&=&\Pr\big\{ L(u_{k}^{2q}[j])=z\big\}\\
&=&\sum_{a,b\,\in\,\supp f_{k+1}^{q}:\; a\llrbox b=z,}f_{k+1}^{q}(a)f_{k+1}^{q}(b).
\end{eqnarray*}
Introduce the notation
\[
f\starbox g(z)=\sum_{a\,\in\,\supp f,\, b\,\in\,\supp g:\; a\llrbox b=z,}f(a)g(b)
\]
and rewrite the recursion as
\begin{equation}
f_{k}^{2q}(z)=f_{k+1}^{q}\starbox f_{k+1}^{q}.\label{eq:f-even}
\end{equation}
While the operation $\starbox$ is not a convolution in the usual sense, we still will use this term.
Now we have the recurrent formulas (\ref{eq:f-odd})
and (\ref{eq:f-even}) which give the required distributions of random variables
$L(u_{0}^{i}[0])$ if the initial probability function
\begin{eqnarray*}
f_{n}^{0}(z) & = & \sum_{b:\ln\frac{W(b|0)}{W(b|1)}=z}W(b\,|\,0).
\end{eqnarray*}
is used as the recursion base.

The probability error $E_i$ is obtained from the probability function $f_0^i$:
\[
E_{i}=\sum_{a\in\supp f_{0}^{i}:\: a<0}f_{0}^{i}(a)+\frac{1}{2}f_{0}^{i}(0).
\]
In what follows we consider a simple case when the convolutions $\star$
and $\starbox$ of two functions are reduced to simple operations on pairs of numbers.

\subsection{The case of binary erasure channel (BEC)}
\begin{figure}[h]
\begin{centering}
\includegraphics[scale=0.6]{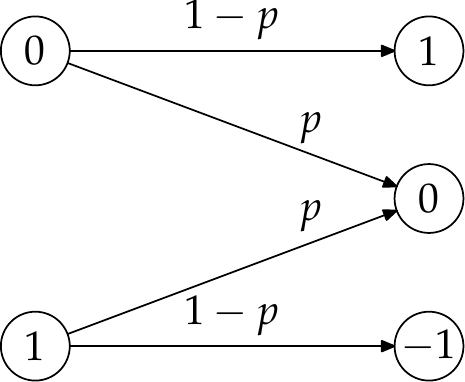}
\par\end{centering}
\caption{Schematic view of binary erasure channel\label{fig:bec}}
\end{figure}

The problem of polar codes construction for BEC was solved in that very article
where the polar codes were introduced, however in a different formulation \cite{DBLP:journals/corr/abs-0807-3917}.
The BEC scheme is shown on fig.~\ref{fig:bec}.

Binary erasure channel is a binary input symmetric channel, possibly simplest one in terms of decoding:
if the received symbol is $1$ or $-1$, the transmitted symbol is unconditionally known.
Initial probability function has the support of two values,
\[
f_{n}^{0}(z)=B_{p}(z)\equiv\begin{cases}
1-p, & z=+\infty,\\
p, & z=0,\\
0, & \mbox{otherwise.}
\end{cases}
\]
The quantity $p\in[0,1]$ is termed the erasure probability.
Consider the convolution of functions $B_p$ and $B_r$ for some $p,r\in[0,1].$
The sum $a+b$ of all possible pairs $(a,b),a\in\supp B_{p},b\in\supp B_{r}$
has the form
\begin{eqnarray*}
0+0 & = & 0,\\
0+(+\infty) & = & +\infty,\\
(+\infty)+0 & = & +\infty,\\
(+\infty)+(+\infty) & = & +\infty.
\end{eqnarray*}
Hence $\supp(B_{p}\star B_{r})=\{0,+\infty\}$. Value of zero
is obtained only with $a=0,b=0$,
\[
(B_{p}\star B_{r})(0)=pr,
\]
therefore
\[
(B_{p}\star B_{r})(+\infty)=1-pr
\]
and
\[
B_{p}\star B_{r}=B_{pr}.
\]
Now we turn to the convolution $B_{p}\starbox B_{r}$. Again consider
all possible pairs $(a,b)$ and the output of $a\llrbox b$:
\begin{eqnarray*}
0\llrbox0 & = & 0,\\
0\llrbox(+\infty) & = & 0,\\
(+\infty)\llrbox0 & = & 0,\\
(+\infty)\llrbox(+\infty) & = & +\infty.
\end{eqnarray*}
Again $\supp(B_{p}\starbox B_{r})=\{0,+\infty\}$. Since the value of
$+\infty$ is obtained only with $a=+\infty,b=+\infty$,
\[
(B_{p}\starbox B_{r})(+\infty)=(1-p)(1-r).
\]
Therefore
\[
(B_{p}\starbox B_{r})(0)=1-(1-p)(1-r)=p+r-pr
\]
and
\[
B_{p}\starbox B_{r}=B_{p+r-pr}.
\]
Thus the formulas (\ref{eq:f-odd}) and (\ref{eq:f-even}) take the form
\begin{eqnarray*}
p_{k}^{2q+1} & = & 2p_{k+1}^{q}-(p_{k+1}^{q})^{2},\\
p_{k}^{2q} & = & (p_{k+1}^{q})^{2},\\
f_{k}^{i} & = & B_{p_{k}^{i}}.
\end{eqnarray*}
The recursion base will be $p_{n}^{0}=p$, the erasure probability of the channel.
It only remains to note that
\[
E_{i}=\frac{1}{2}p_{0}^{i},\quad i=\overline{0,2^{n}-1}.
\]
The case of general symmetric channel is considered in the next subsection.

\subsection{The general case}
In the general case the complexity of exact computation of convolutions becomes
too high since the support cardinality (and memory requirements)
for the probability function grows exponentially
with the code length. Approximation of the probability functions using
a uniform grid is quite natural. Denote by $\delta$ the grid step and by
$\Omega_i$ the grid cell number $i$ with $i=\overline{-Q,Q}$: 
\begin{eqnarray*}
\Omega_{0} & = & \left(-\frac{\delta}{2},\,\frac{\delta}{2}\right),\\
\Omega_{i} & = & \left[i\delta-\frac{\delta}{2},\, i\delta+\frac{\delta}{2}\right),\quad i=\overline{1,Q-1},\\
\Omega_{i} & = & \left(i\delta-\frac{\delta}{2},\, i\delta+\frac{\delta}{2}\right],\quad i=\overline{-Q+1,-1},\\
\Omega_{Q} & = & \left[Q\delta-\frac{\delta}{2},\,+\infty\right),\\
\Omega_{-Q} & = & \left(-\infty,\,-Q\delta+\frac{\delta}{2}\right],
\end{eqnarray*}
where $Q$ is a positive integer which will be called \emph{the number of quantization
levels.} Thus the grid consists of $2Q+1$ cells. The points $i\delta$ will be called the grid nodes.
All cells except for extreme ones have grid nodes as centers.

Define the grid projection operator. Let $f$ be some probability function. For each grid node,
sum all nonzero values of $f$ which belong to the corresponding grid cell%
\footnote{if $f$ is the probability density function, let $\hat{f}(i\delta)=\int_{\Omega_{i}}f(\omega)d\omega$%
}:
\begin{equation}
\hat{f}(i\delta)=\sum_{z\,\in\,\Omega_{i}\cap\supp f}f(z).\label{eq:grid-projection}
\end{equation}
Let $g$ and $h$ be two functions supported at grid nodes. The convolutions
$g\star h$ and $g\starbox h$ can take nonzero values outside the set of grid nodes.
Use the projection  (\ref{eq:grid-projection}) to restrict the resulting function to the grid.

Note that the convolution $g\star h$ can have nonzeros only at points
$i\delta$ with $i$ from $-2Q$ up to $2Q$.
Among these points, only those with  $|i|>Q$ are not grid nodes. If $i>Q$, these points
belong to the rightmost cell  $\Omega_{Q}$,
and if $i<-Q$, to the leftmost cell $\Omega_{-Q}$.
Thus the projection operation for the convolution result consists in summing
the values outside the interval  $[-Q\delta,Q\delta]$.
The convolution  $g\starbox h$, on the contrary, is not supported outside
the interval  $[-Q\delta,Q\delta]$ because  $|a\llrbox b|\leq\min(|a|,|b|)$.

Denote by $\mathrm{nearest}(x)$ the index of the cell $\Omega_{i}$ containing the point $x$.
In other words, $\mathrm{nearest}(x)$ is the index of the grid node closest to $x$.
Then the approximate computation of convolutions  $\star$ and $\starbox$
described above corresponds to algorithms  \ref{alg:star} and \ref{alg:starbox}, respectively.

\begin{algorithm}
\caption{Computation of the projection $\hat{f}$ of $g\star h$ with arguments supported
at grid nodes\label{alg:star}}
\begin{algorithmic}[1]
\State $f \gets g \star h$
\State $\hat{f}(x) \gets 0 \quad \forall x$
\For {$i \gets -Q, \ldots, Q$}
\State $\hat{f}(i\delta) \gets f(i\delta)$
\EndFor
\For {$i \gets Q+1, \ldots, 2Q$}
	\State $\hat{f}(Q\delta) \gets \hat{f}(Q\delta) + f(i\delta)$
	\State $\hat{f}(-Q\delta) \gets \hat{f}(-Q\delta) + f(-i\delta)$
\EndFor
\end{algorithmic}
\end{algorithm}

\begin{algorithm}
\caption{Computation of the projection $\hat{f}$ of $g\starbox h$ with arguments supported
at grid nodes\label{alg:starbox}}
\begin{algorithmic}[1]
\State $\hat{f}(x) \gets 0 \quad \forall x$
\For {$i \gets -Q, \ldots, Q$}
\For {$j \gets -Q, \ldots, Q$}
	\State $k \gets \mathrm{nearest}(i\delta \llrbox j\delta) $
	\State $\hat{f}(k\delta) \gets
			\hat{f}(k\delta) + g(i\delta)h(j\delta)$
\EndFor
\EndFor
\end{algorithmic}
\end{algorithm}

Since the grid is uniform, the convolution
$f\gets g\star h$
from the first step of the algorithm  \ref{alg:star} may be computed
in $\mathcal{O}(Q\log_{2}Q)$ operations using FFT. The rest of the algorithm takes only
$\mathcal{O}(Q)$ operations, hence the overall complexity of the algorithm \ref{alg:star} is
 $\mathcal{O}(Q\log_{2}Q)$ operations.
The complexity of the algorithm \ref{alg:starbox} is $\mathcal{O}(Q^{2})$,
which is much worse.

Convolutions $\star$ and $\starbox$ arise also in the problem of optimizing
the weight distributions for rows and columns of the LDPC check matrix.
Results from this area may be used for the design of fast version of the algorithm \ref{alg:starbox},
namely the algorithm from  \cite{Richardson:2008:MCT:1795974}.
It is based on the following inequalities for the quantity $k$ appearing in the line $4$
of the algorithm  \ref{alg:starbox},
\[
\min(|i|,|j|)-\left(\frac{\ln2}{\delta}-\frac{1}{2}\right)<|k|<\min(|i|,|j|).
\]
Also, $\sgn k=\sgn i\cdot\sgn j$, i.e. the quantity $\sgn i\cdot\sgn j\cdot\min(|i|,|j|)$
estimates $k$ with an error not exceeding
$M(\delta)=\lceil\frac{\ln2}{\delta}-\frac{1}{2}\rceil$.
This observation helps to reduce the complexity of the algorithm to
$\mathcal{O}(Q\cdot M(\delta))$ operations.
However taking finer grid makes  $M(\delta)$ larger, and the speedup smaller.
However the speedup is noticeable. Let $A=\delta Q$ be the rightmost grid node
and $[-A,A]$ the segment containing all grid nodes. Typical values used
in our numerical experiments were $A=60,$ $Q=2^{13}$.
In this case $\delta=\frac{A}{Q}\approx0,007324$ and $M(\delta)=95$.

Thus making the grid projection of the initial probability function  $f_{n}^{0}$
and substituting approximations for the exact computations which use formulas  (\ref{eq:f-odd})
and  (\ref{eq:f-even}) we obtain a numerical method for computation of probability
errors $E_{i}$. While the accuracy analysis for this method remains an open question,
our numerical experiments show that good accuracy can be achieved without refining the grid too much.

\subsection{Performance analysis}

Construction procedure described above implies that the polar code is built
for a concrete channel. In practice, channel properties may change with time,
therefore it is important to analyze the performance of the constructed code
for channel models with different noise levels. For most modern coding systems,
in particular for low-density parity check codes, the only available tool
is the Monte-Carlo simulation.

For polar codes, such analysis is available in much less expensive way.
To obtain the upper bound for the block error probability, one can compute
the error probabilities $E_i$ by the method used in code construction
and sum these quantities over indices of information bits. Numerical experiments show
that this estimate is quite accurate.

\subsection{Numerical experiments}

It is instructive to check the quality of the estimates given by the described
performance analysis method. To this end, one can compare the Monte-Carlo
simulation results and the obtained estimate for some concrete code.
Using random number generator, form ``received'' vector $y$ satisfying
the channel model and decode it. For large number of trials $N_T$,
the decoder will make $N_E$ errors. We can estimate the block error probability
as follows,
\[
P_{E}\approx\frac{N_{E}}{N_{T}}.
\]
According to the central limit theorem, with the probability of some $95\%$
this estimate belongs to the confidence interval of the radius
\begin{equation}
1,96\sqrt{\frac{\sigma^{2}}{N_{T}}},\label{eq:confidence}
\end{equation}
where $\sigma^{2}$ is the variance of the random variable
taking the value of $1$ if the decoder makes an error and $0$ otherwise
 \cite{Rubino:2009:RES:1643623}.
Exact value of $\sigma^{2}$ is
\[
\sigma^{2}=P_{E}(1-P_{E})
\]
and while it is unknown, we can estimate it using the sample variance formula
\[
\sigma^{2}\approx\frac{N_{T}}{N_{T}-1}\cdot\frac{N_{T}}{N_{E}}\left(1-\frac{N_{T}}{N_{E}}\right)
\]
Thus the Monte-Carlo method has the accuracy of the order
$N_{T}^{-\frac{1}{2}}$
which implies large computational costs. For  $P_{E}\ll1$,
obtaining a $50\%$ confidence estimate requires according to formula (\ref{eq:confidence}),
some  $4\cdot1,96^{2}\cdot P_{E}^{-1}$
trials. For example, if $P_{E}=10^{-7}$, one will need $15\cdot10^{7}$ trials.
Further, with $10\%$ confidence level the number of trials increases up to  $100\cdot1,96^{2}\cdot P_{E}^{-1}$.
If $P_{E}=10^{-7}$, this number will be approximately $3,8\cdot10^{9}$.
Therefore in Monte-Carlo simulations we restrict the noise level to interval
corresponding to $P_E$ exceeding $10^{-5}$.

\begin{figure}
\begin{centering}
\includegraphics[scale=0.7]{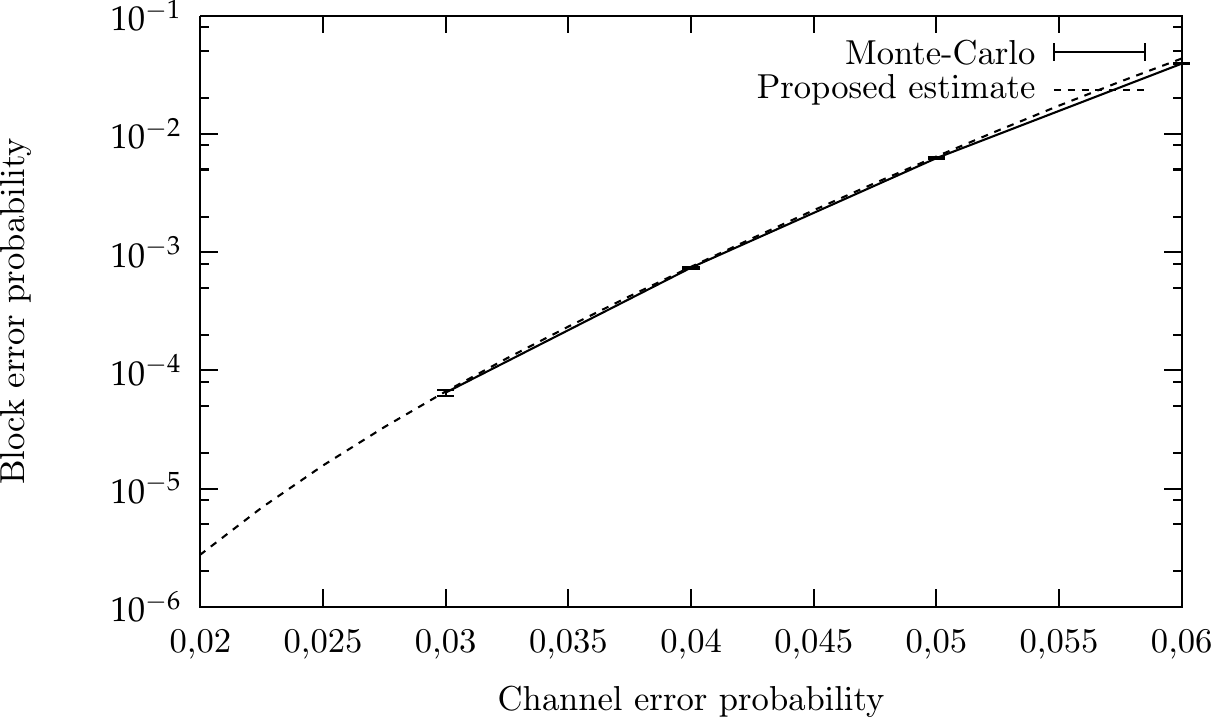}
\par\end{centering}

\caption{Performance of the polar code of length $1024$, rate $\frac{1}{2}$
on binary symmetric channel estimated by Monte-Carlo simulations and proposed
analysis method\label{fig:mc-vs-a}}
\end{figure}
For this experiment we constructed a polar code of length $1024$ and rate
$\frac{1}{2}$ for binary symmetric channel with error probability $0.06$.
Monte-Carlo simulation was run for binary symmetric channels with various
error probabilities. Also, an estimate of block error probability
was computed using the proposed analysis method. Obtained graphs are shown on fig.~\ref{fig:mc-vs-a}.
One can see that the results produced independently in two different ways
are very close.

\begin{figure}
\begin{centering}
\includegraphics[scale=0.7]{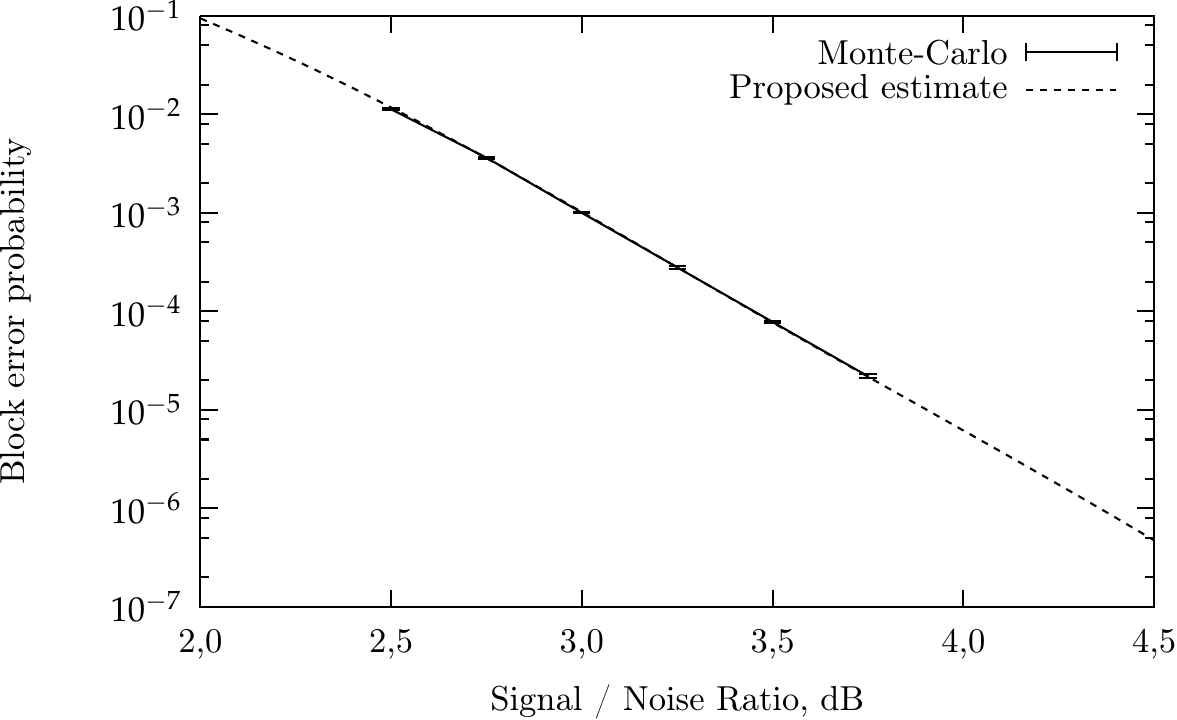}
\par\end{centering}

\caption{Performance of the polar code of length $1024$, rate $\frac{1}{2}$
on AWGN channel estimated by Monte-Carlo simulations and proposed
analysis method\label{fig:1024-awgn}}
\end{figure}

Consider now a different channel model, an AWGN channel with binary input
and additive normal noise. Output alphabet for this channel is the real axis,
while the transition function has the form
\[
W(y\,|\, x)=\frac{1}{\sqrt{2\pi\sigma^{2}}}\exp\left(-\frac{(y-(1-2x))^{2}}{2\sigma^{2}}\right),\quad x\in\{0,1\}.
\]
In other words, transmission over such channel amounts
to mapping input bits $0$ and $1$ to symbols $1$ and $-1$, respectively,
and adding afterwards normal noise with zero average and variance
$\sigma^{2}$. 
Note that this channel has continuous output alphabet and does not fit to
previous sections theory.
However a similar numerical experiment is perfectly possible for a
discrete approximation of this channel.
Instead of $\sigma^{2}$, on the horizontal axis we plot the signal/noise ratio
in decibels
\[
\mbox{SNR (dB)}=10\log_{10}\frac{1}{\sigma^{2}}.
\]
We constructed a polar code of length $1024$ and rate $\frac{1}{2}$
for the noise level $3$dB. Monte-Carlo simulation was run for various
noise levels. Also, an estimate was computed using the proposed analysis method.
The results are shown in fog.~\ref{fig:1024-awgn}. One can see that
the graphs again are almost identical.

\begin{figure}
\begin{centering}
\includegraphics[scale=0.7]{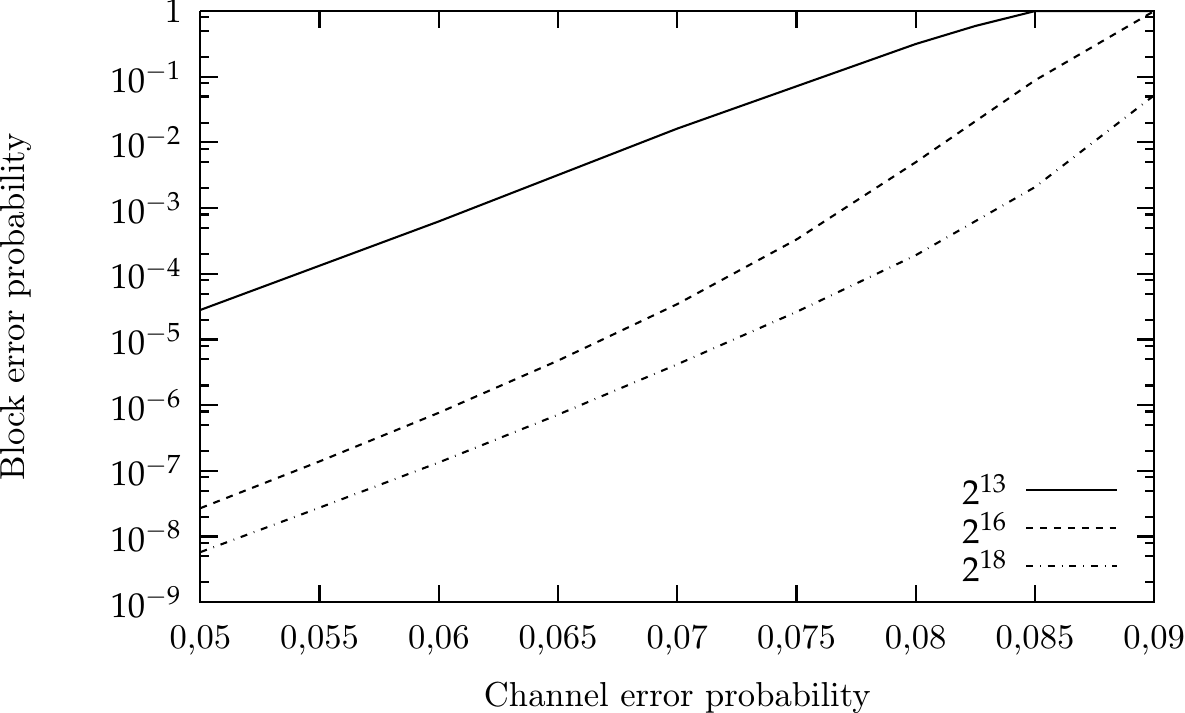}
\par\end{centering}

\caption{Performance of polar codes of rate $\frac{1}{2}$ and different lengths
for binary symmetric channel\label{fig:polar-bsc}}
\end{figure}

Fig.~\ref{fig:polar-bsc} shows performance graphs for polar codes of rate $\frac12$ and of lengths
 $2^{13}=8192$, $2^{16}=65536$ and $2^{18}=262144$
for binary symmetric channel. For code rate $\frac12$ and binary symmetric channel,
the Shannon limit corresponds to $p=0.11$. One can see that the convergence to Shannon limit
is rather slow. Next section is devoted to generalization of polar codes which
allows to increase the convergence rate.

\section{Polarization kernels\label{sec:kernels}}
For the definition of polar codes, the following matrix was used in section~\ref{sec:polar},
\[
G_{2}=\left[\begin{array}{cc}
1 & 0\\
1 & 1
\end{array}\right].
\]
One can use another invertible matrix $G$ of arbitrary order $l$.
Then the hierarchical graph construction will involve taking $l$ copies
of encoder graph, instead of doubling.
New variables will be expressed in terms of old ones by the formula
\begin{eqnarray*}
&&[u_{k+1}^{i}[lj],u_{k+1}^{i}[lj+1],\ldots,u_{k+1}^{i}[lj+l-1]=\\
&&\qquad[u_{k}^{li}[j],u_{k}^{li+1}[j],\ldots,u_{k}^{li+l-1}]\cdot G,
\end{eqnarray*}
and old variables in terms of new ones, by the formula 
\begin{eqnarray*}
&&\qquad[u_{k}^{li}[j],u_{k}^{li+1}[j],\ldots,u_{k}^{li+l-1}]=\\
&&[u_{k+1}^{i}[lj],u_{k+1}^{i}[lj+1],\ldots,u_{k+1}^{i}[lj+l-1]]\cdot G^{-1}.
\end{eqnarray*}
After $n$ steps of graph construction, the code length will be $l^n$
and the decoder graph will consist of $n$ layers, each having $J_{k}=l^{k}$
groups of $S_{k}=l^{n-k}$ variables. The vector of output symbols $Y_{k}[j]$
corresponding to group $j$ of layer $k$, still will consist of contiguous bits
from index  $j\cdot S_{k}$ up to index  $(j+1)\cdot S_{k}-1$ inclusively.
The problem of computation of quantities
\[
L(u_{k}^{li}[j]),L(u_{k}^{li+1}[j]),\ldots,L(u_{k}^{li+l-1}[j])
\]
using the values of
\[
L(u_{k+1}^{i}[lj]),L(u_{k+1}^{i}[lj+1]),\ldots,L(u_{k+1}^{i}[lj+l-1])
\]
leads to a graph analogous to the shown in the fig.~\ref{fig:simplified},
this time isomorphic to the Tanner graph of the matrix  $G^{-1}$.
In general this problem cannot be reduced to belief propagation algorithm
working on a tree and requires exponential in $l$ number of operations.
Computation of  $L(u_{k}^{li+m}[j])$ is always possible by enumeration
of all possible events. For brevity, denote
$u_{m}=u_{k}^{li+m}[j],$$x_{m}=u_{k+1}^{i}[lj+m],Y=Y_{k}[j]$,$x=[x_{0},x_{1},\ldots,x_{l-1}]$.
Then
\begin{equation}
L(u_{m})=\ln\frac{\Pr\{Y\,|\, u_{m}=0;u_{0},u_{1},\ldots,u_{m-1}\}}{\Pr\{Y\,|\, u_{m}=1;u_{0},u_{1},\ldots,u_{m-1}\}}.\label{eq:l-u-m}
\end{equation}
Let $X_{a},a=0,1$ be the set of all vectors $x$ such that
\[
[u_{0},u_{1},\ldots,u_{m-1},a,***]=xG^{-1},
\]
where $***$ stands for $l-m-1$ arbitrary bits. Then
\begin{eqnarray*}
\Pr\{Y&|& u_{m}=a;u_{0},u_{1},\ldots,u_{m-1}\} =\\
 && \frac{1}{|X_{a}|}\sum_{x\in X_{a}}\Pr\{Y\,|\, x\}=\\
 && \frac{1}{|X_{a}|}\sum_{x\in X_{a}}\prod_{t=0}^{l-1}\Pr\{Y^{t}\,|\, x_{t}\},
\end{eqnarray*}
where $Y^{t}$ is component $t$ of $Y$. Inserting the last equality
for $a=0,1$ into numerator and denominator of (\ref{eq:l-u-m}),
we get
\[
L(u_{m})=\ln\frac{\sum_{x\in X_{0}}\prod_{t=0}^{l-1}\Pr\{Y^{t}\,|\, x_{t}\}}{\sum_{x\in X_{1}}\prod_{t=0}^{l-1}\Pr\{Y^{t}\,|\, x_{t}\}}.
\]
Dividing the numerator and denominator by
$\prod_{t=0}^{l-1}\Pr\{Y^{t}\,|\, x_{t}=1\},$
we get
\[
L(u_{m})=\ln\frac{\sum_{x\in X_{0}}\prod_{t=0}^{l-1}l(x_{t})^{1\oplus x_{t}}}{\sum_{x\in X_{1}}\prod_{t=0}^{l-1}l(x_{t})^{1\oplus x_{t}}},
\]
where
\[
l(x_{t})=\frac{\Pr\{Y^{t}|x_{t}=0\}}{\Pr\{Y^{t}|x_{t}=1\}}=e^{L(x_{t})}.
\]
Using the last equality, write
\begin{equation}
L(u_{m})=\ln\frac{\sum_{x\in X_{0}}\exp(\sum_{t=0}^{l-1}(1\oplus x_{t})L(x_{t}))}{\sum_{x\in X_{1}}\exp(\sum_{t=0}^{l-1}(1\oplus x_{t})L(x_{t}))}.\label{eq:kernel-recursion}
\end{equation}
This gives the recursive formula for computation of
$L(u_{k}^{li+m}[j])$ in the case of arbitrary matrix $G$,
however involving sums with exponential in $l$ number of terms.

\subsection{Obtaining the recursive formulas}
\begin{figure}[b]
\begin{centering}
\includegraphics[scale=0.8]{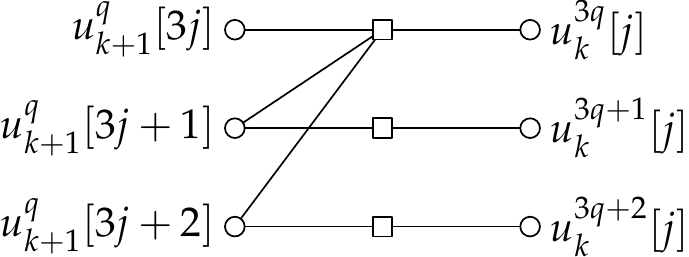}
\par\end{centering}

\caption{Graph for the recurrent relations for the matrix $G_{3}$\label{fig:g3}}
\end{figure}

For some polarization kernels, the formulas (\ref{eq:kernel-recursion})
may be replaced by simpler relations containing familiar operations
$+$  and $\llrbox$. For example, consider the matrix
\[
G_{3}=\left[\begin{array}{ccc}
1 & 0 & 0\\
1 & 1 & 0\\
1 & 0 & 1
\end{array}\right].
\]
Note that $G_{3}^{-1}=G_{3}$ and draw the Tanner graph analogous to the shown in
the fig.~\ref{fig:simplified} but corresponding to $G_{3}^{-1}$
(fig.~\ref{fig:g3}).

\begin{figure}
\begin{centering}
\includegraphics[scale=0.8]{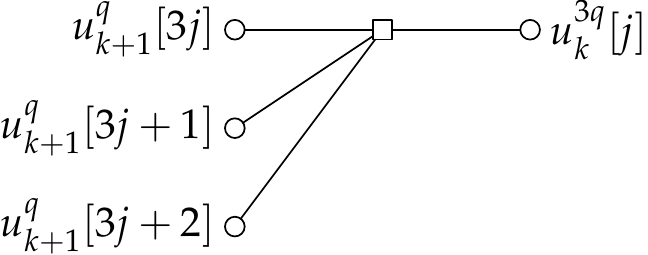}
\par\end{centering}

\caption{Graph from fig.~\ref{fig:g3} after removal of excessive vertices\label{fig:g3-1}}
\end{figure}

We now find the expression for $L(u_{k}^{3q}[j])$. 
The vertices $u_{k}^{3q+1}[j]$ и $u_{k}^{3q+2}[j]$ may be removed from the graph
together with incident equations.
We obtain the graph shown in the fig.~\ref{fig:g3-1}.
Using results of section~ \ref{sec:bp-on-trees}, one can write
\[
L(u_{k}^{3q}[j])=L(u_{k+1}^{q}[3j])\llrbox L(u_{k+1}^{q}[3j+1])\llrbox L(u_{k+1}^{q}[3j+2]).
\]

\begin{figure}
\begin{centering}
\includegraphics[scale=0.8]{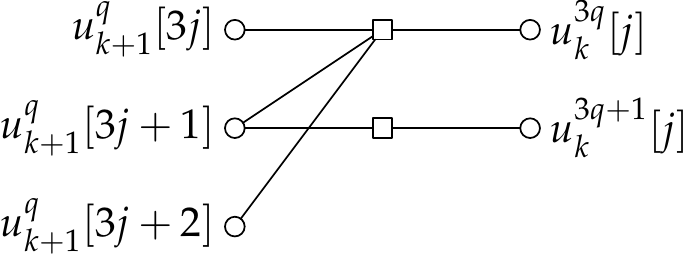}
\par\end{centering}
\caption{Graph from fig.~\ref{fig:g3} after removal of excessive vertex\label{fig:g3-2}}
\end{figure}

In order to find $L(u_{k}^{3q+1}[j])$, the quantity  $u_{k}^{3q}[j]$
should be considered known exactly. We have to remove the vertex $u_{k}^{3q+2}[j]$
from the graph in fig.~\ref{fig:g3} and the incident equation (fig.~\ref{fig:g3-2}).
From the last graph we conclude that
\begin{eqnarray*}
&&L(u_{k}^{3q+1}[j])= L(u_{k+1}^{q}[3j+1])+\\
&&\qquad(-1)^{u_{k}^{3q}[j]}(L(u_{k+1}^{q}[3j])\llrbox L(u_{k+1}^{q}[3j+2])).
\end{eqnarray*}
Similarly we can write the third formula:
\begin{eqnarray*}
&&L(u_{k}^{3q+2}[j])=L(u_{k+1}^{q}[3j+2])+\\
&&\qquad(-1)^{u_{k}^{3q}[j]\oplus u_{k}^{3q+1}[j]}L(u_{k+1}^{q}[3j]).
\end{eqnarray*}

In an analogous way, one can try to obtain recurrent formulas for an arbitrary
polarization kernel. If the Tanner graph for some fixed index is not tree-like,
one can try to amend it by adding some equations to other ones.
Further, if a cycle contains an exactly known bit, the cycle can be broken
by doubling the respective vertex.
Unfortunately, starting from $l=5$ the tree-like graph
can be obtained only for some of polarization kernels.

Those kernels which admit simple formulas also admit simple code construction
and analysis. For instance, in the example considered above the recurrent formulas for probability
functions have the form
\begin{eqnarray*}
f_{k}^{3q} & = & f_{k+1}^{q}\starbox f_{k+1}^{q}\starbox f_{k+1}^{q},\\
f_{k}^{3q+1} & = & f_{k+1}^{q}\star(f_{k+1}^{q}\starbox f_{k+1}^{q}),\\
f_{k}^{3q+2} & = & f_{k+1}^{q}\star f_{k+1}^{q}.
\end{eqnarray*}
For polarization kernels which do not admit simple recurrent formulas,
the problem of code construction is open. In general, the computation
of probability functions  $f_{k}^{lq+j}$ via $f_{k+1}^{q}$ is a multidimensional
summation problem. Possible solutions are Monte-Carlo method and approximations
by normal distribution.

\section{Concatenated polar codes\label{sec:new-polar}}

In this section we consider a method of performance improvement for polar codes
in which short classic error correcting codes are used together with polar codes.

Let $C_{1},C_{2},\ldots,C_{q}$ be a set of linear codes of equal length $M$.
Let $K_{i}$ be the number of information bits in the code $C_{i}$.
Let $V$ be some $M\times N$ matrix each of whose elements
is $0$ or $1$. Denote by $v_{ji}$ with $0\le i<N$ and $0\le j<M$
the elements of $V$, by $v^j$ its row $j$
and by $v_i$ its column $i$. For all $i=\overline{0,N-1}$
choose some integer $a_i$ in the range $1$ to $q$. We consider only such matrices $V$
whose columns $v_i$ are codewords of $C_{a_i}$, i.e.
\begin{equation}
v_{i}\in C_{a_{i}},\quad i=\overline{0,N-1}.\label{eq:constraint}
\end{equation}

Consider an arbitrary polar code of length $N$ and rate $1$, i.e. without redundancy,
with matrix generator $G\in\GF(2)^{N\times N}$.
Encode each row of $V$ with this polar code obtaining a new matrix
$X\in\GF(2)^{M\times N}$:
\begin{equation}
X=VG.\label{eq:mtx}
\end{equation}
If the matrix $X$ is ``reshaped'' into a row, one can consider the set of all such
possible rows subject to restriction (\ref{eq:constraint})
as a linear code of length  $M\cdot N$ and rate
\[
\frac{K}{MN}=\frac{1}{MN}\sum_{i=0}^{N-1}K_{a_{i}}.
\]
Thus obtained linear code we will call the \emph{concatenated polar code}.
Let $Y$ be the matrix received after the transmission of $X$ through the channel
and let $y^j$ be its row $j$. The decoder works by applying alternatively
the steps of successive cancellation
method for rows of $Y$ and maximum likelihood decoder for its columns.

In order to decode the column $v_0$, compute for each row of $Y$
independently the logarithmic likelihood ratios
\[
L(v_{j,0})=\ln\frac{\Pr\{y^{j}|v_{j,0}=0\}}{\Pr\{y^{j}|v_{j,0}=1\}},
\]
just like in the usual successive cancellation method.
Then the values $L(v_{j,0})$, $j=\overline{0,M-1}$ gathered in a vector $y$ are
given as input to ML-decoder for the code $C_{a_0}$. The most likely codeword
$w\in C_{a_{0}}$ produced on output is taken as an estimate of $v_0$.

Next we compute the estimate of $v_1$. Assuming $v_0$ already known,
again compute for each row independently the LLRs
\[
L(v_{j,1})=\frac{\Pr\{y^{j}|v_{j,1}=0;\, v_{j,0}\}}{\Pr\{y^{j}|v_{j,1}=1;\, v_{j,0}\}},
\]
concatenate the values $L(v_{j,1})$ into a vector $y$, which will be the input of
ML-decoder for the code $C_{a_1}$. The obtained codeword is taken as an estimate
of $v_1$. Next, assuming $v_0$ and $v_1$ exactly known, compute the estimate of $v_2$
etc.

Note that the polar codes are a special case of concatenated polar codes
for $M=1$, $q=2$ and $C_{1}=\{0\},$ $C_{2}=\{0,1\}$. In this case, 
bit $i$ is frozen if $a_{i}=1$, and it is information bit, if $a_{i}=2$.

\subsection{Code construction and analysis}
Let $E_i$ be the error probability for estimation of column $i$
under the constraint that all previous columns were estimated error-free.
Write again the upper bound for block error probability:
\begin{equation}
P_{E}\leq\sum_{i=0}^{N-1}E_{i}.\label{eq:upper-bound}
\end{equation}

Fix some symmetric channel $W$, set of codes $C_{1},\ldots,C_{q}$
of length $M$, polar code of length $N$ and rate $1$.
We require to construct a concatenated polar code of given rate $k/N$,
i.e. choose numbers
$a_{0},a_{1},\ldots,a_{N-1}$ such that
\begin{equation}
\sum_{i=0}^{N-1}K_{a_{i}}=K.\label{eq:sum-k-equals-k}
\end{equation}
We will choose these numbers so as to minimize the upper bound (\ref{eq:upper-bound}).
Denote by $E_{i}^{k}$ the error probability for estimation of the column
$v_{i}$ under the constraint that all previous columns were estimated error-free
and $a_{i}=k$. Note that $E_{i}^{k}$ does not depend on $a_{j}$ for all
$j\neq i$. For a concrete choice of $a_{0},a_{1},\ldots,a_{N-1}$
we can write the following upper bound for $P_{E}$,
\begin{equation}
P_{E}\leq\sum_{i=0}^{N-1}E_{i}^{a_{i}}.\label{eq:upper-bound-re}
\end{equation}
Assume for now that for all $i=\overline{0,N-1}$ and $k=\overline{1,q}$
we can compute $E_{i}^{k}$. In order to choose the optimal $a_{0},a_{1},\ldots,a_{N-1}$,
we will use the dynamic programming method. Let
 $F(s,t),s=\overline{0,N-1},t=\overline{0,K}$ be the minimal possible value of the sum
\[
\sum_{i=0}^{s}E_{i}^{a_{i}}
\]
under the constraint
\begin{equation}
\sum_{i=0}^{s}K_{a_{i}}=t,\label{eq:sum-k-equals-t}
\end{equation}
or let $F(s,t)=+\infty$, if there is no sets of $a_{i}$ satisfying (\ref{eq:sum-k-equals-t}).
For convenience, set $F(s,t)=+\infty$ for $t<0$. It is easy to note that
\begin{equation}
F(s,t)=\min_{l}\left(F(s-1,t-K_{l})+E_{t}^{l}\right).\label{eq:f-s-plus-one}
\end{equation}
Introduce the notation%
\footnote{if the minimum is attained for several values of $l$, any of those can serve
as $A(s,t)$.}
\[
A(s,t)=\arg\min_{l}\left(F(s-1,t-K_{l})+E_{t}^{l}\right).
\]
To make the formula (\ref{eq:f-s-plus-one}) correct also for $s=0$, let
\begin{eqnarray*}
F(-1,0) & = & 0,\\
F(-1,t) & = & +\infty,\quad t\neq0.
\end{eqnarray*}
Now using (\ref{eq:f-s-plus-one}) for sequential computation of $F(s,t)$ for
$s=0,1,2,\ldots$ and all $t$, and saving the corresponding quantities $A(s,t)$, 
one can compute $F(N-1,K),$ which by definition is the minimal possible value of the sum
(\ref{eq:upper-bound-re}) under the constraint (\ref{eq:sum-k-equals-k}).
If $F(N-1,K)=+\infty$, there is no set of $a_{0},a_{1},\ldots,a_{N-1}$
satisfying the constraint (\ref{eq:sum-k-equals-k}).

Let $F(N-1,K)\neq+\infty$. In order to recover the sequence $a_{0},a_{1},\ldots,a_{N-1}$
delivering the minimum to the sum (\ref{eq:upper-bound-re}), we make a pass in reverse order
utilizing the saved quantities
$A(s,t)$,
\begin{eqnarray*}
a_{N-1} & = & A(N-1,K),\\
a_{N-2} & = & A(N-2,K-K_{a_{N-1}}),\\
a_{N-3} & = & A(N-3,K-K_{a_{N-1}}-K_{a_{N-2}}),\\
 & \vdots\\
a_{i} & = & A(i,K-\sum_{l=i+1}^{N-1}K_{a_{l}}),\\
 & \vdots\\
a_{0} & = & A(0,K-\sum_{l=1}^{N-1}K_{a_{l}}).
\end{eqnarray*}

Now we get back to the problem of estimating $E_{i}^{k}$. Since the channel is symmetric
and the code is linear, we assume the all-zero codeword is sent.
Suppose that the columns $v_{0},v_{1},\ldots,v_{i-1}$ have been estimated correctly
and the decoder is to estimate $v_i$. Next the ML-decoder for the code $C_k$
takes on input the vector
\[
\lambda=[L(v_{0,i}),L(v_{1,i}),L(v_{2,i}),\ldots,L(v_{M-1,i})].
\]
For convenience, introduce the notation $\lambda_{j}\equiv L(v_{j,i})$. The components
of $\lambda$ are i.i.d. random variables.
Their probability function (or pdf) $f_{i}$
can be computed approximately using the method described in section~\ref{sec:design}.
We can assume that the column $v_i$ is transmitted via some symmetric channel
with LLR distribution
$f_{i}$.
Thus the problem of computing $E_{i}^{k}$ is reduced to the estimation
of error probability for the ML-decoder on a channel
with given probability function $f_i$.
As stated in the introduction, the ML-decoder minimizes the linear functional
\[
\phi(c)=\sum_{j=0}^{M-1}c_{j}\lambda_{j},
\]
where $c=[c_{0},c_{1},\ldots,c_{M-1}]$ runs over all codewords of the code
$C_{k}$. For the all-zero codeword the functional $\phi$ is zero.
Hence if the decoding error occurs, there necessarily exists some
codeword $c'$ such that $\phi(c')\leq0$. The last inequality can be rewritten
as the sum of $w_{H}(c')$ terms,
\[
\sum_{j\,\in\,\supp c'}\lambda_{j}\leq0.
\]
Some nonzero codeword $c'$ will be strictly more preferable than $0$
if $\phi(c')<0$ and in this case the decoder error will surely occur.
If $\phi(c')=0$, the decoder may choose the correct codeword among those
which zero the functional $\phi$. For simplicity assume that $\phi(c')=0$
also implies the decoder error.
Write the probability of the event that for a fixed $c'$
the inequality $\phi(c')\leq0$ holds as
\[
\Pr\left\{ \sum_{j\,\in\,\supp c'}\lambda_{j}\leq0\right\} .
\]
The sum consists of $w_{H}(c')$ i.i.d. random variables with the probability function
$f_{i}$, therefore the probability function of the sum is
\[
f_{i}^{\star w_{H}(c')}\equiv\underbrace{f_{i}\star f_{i}\star\ldots\star f_{i}}_{w_{H}(c')\mbox{ times}}
\]
It follows that the probability of the event
$\phi(c')\leq0$ depends only on the weight $w$ of the codeword $c'$ and it can be written as
\[
P(f,w)=\sum_{x\in\supp f^{\star w}:\: x\leq0}f^{\star w}(x).
\]
The main contribution in the error probability is made by codewords of minimal weight.
Let $d_{k}$ be the code distance of the code $C_{k}$, and let $m_{k}$ be the number of
different codewords of weight $d_{k}$ in the code $C_{k}$.
Then the probability $E_{i}^{k}$ may be estimated as
\begin{equation}
E_{i}^{k}\approx m_{k}\cdot P(f_{i},d_{k}).\label{eq:estimate}
\end{equation}
Experiments show that this value is likely to overestimate the real error
probability (computed by a Monte-Carlo simulation) by a constant factor which does not depend
on the channel. For this reason in experiments reported in this paper
a simple empiric technique was used to correct the multiplier $m_k$.
Each of the codes  $C_{1},C_{1},\ldots,C_{q}$
was simulated on an AWGN channel with different SNR ratios to obtain
its FER curve. The number $m_k$ was chosen so that the estimate (\ref{eq:estimate})
fitted the experimental curve best. We do not have a theoretical justification of this procedure,
however the results of numerical experiments show its high accuracy.

In a similar way one can estimate the FER of a concrete concatenated polar code
on a given channel. It is sufficient to approximate numerically the sum
\begin{equation}
\sum_{i=0}^{N-1}E_{i}^{a_{i}}\label{eq:fer-bound}
\end{equation}
and take it as an upper bound for block error rate.

\subsection{Numerical experiments}
For the construction of concatenated polar codes consider a set of $26$ different
codes of length $32$. Numbers of information bits $K_i$ and code distances $d_i$
for each code are given in the table~\ref{tab:palette}.

\begin{figure}[h!]
\begin{centering}
\includegraphics[scale=0.7]{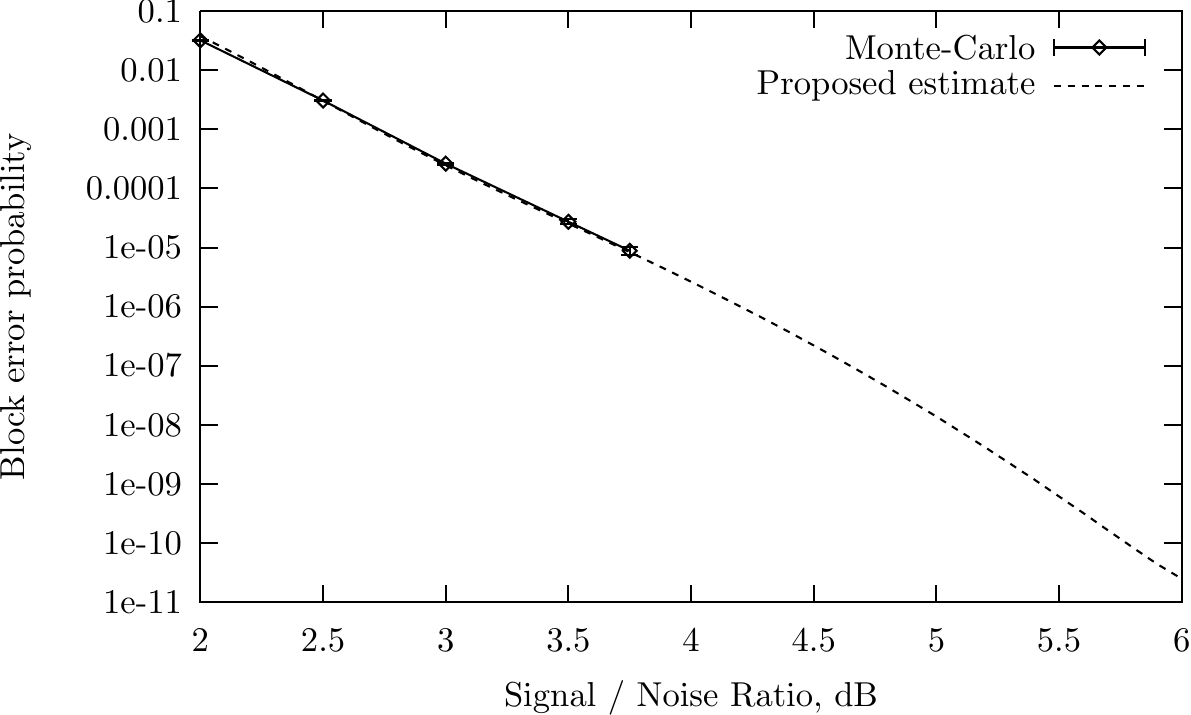}
\par\end{centering}

\caption{Performance of the concatenated polar code of length $1024=32\times32$ and rate
$\frac{1}{2}$ on an AWGN channel estimated
using Monte-Carlo method and using the proposed method\label{fig:ecc-mc-vs-est}}
\end{figure}

The first interesting question is the accuracy of the block error estimate
 (\ref{eq:fer-bound}) which is computed approximately.
In the fig.~$(\ref{fig:ecc-mc-vs-est})$ we show the performance graph
of the concatenated polar code of length $1024$ and rate $\frac12$ on an AWGN channel.
The code was constructed for the channel with SNR$=2.5$dB.
The solid curve represents the Monte-Carlo estimate, the dotted curve
is the estimate (\ref{eq:fer-bound}) computed approximately.
One can see that the curves are practically identical within the limit
of applicability of the Monte-Carlo method.

\begin{table}
\begin{centering}
{\tiny
\begin{tabular}{c|cccc|cccc|cccc|cc}
$i$ & $K_{i}$ & $d_{i}$ &  & $i$ & $K_{i}$ & $d_{i}$ &  & $i$ & $K_{i}$ & $d_{i}$ &  & $i$ & $K_{i}$ & $d_{i}$\tabularnewline
\cline{1-3} \cline{5-7} \cline{9-11} \cline{13-15} 
1 & 0 & $\infty$ &  & 8 & 7 & 14 &  & 15 & 21 & 6 &  & 22 & 28 & 2\tabularnewline
\cline{1-3} \cline{5-7} \cline{9-11} \cline{13-15} 
2 & 1 & 32 &  & 9 & 8 & 13 &  & 16 & 22 & 5 &  & 23 & 29 & 2\tabularnewline
\cline{1-3} \cline{5-7} \cline{9-11} \cline{13-15} 
3 & 2 & 21 &  & 10 & 11 & 12 &  & 17 & 23 & 4 &  & 24 & 30 & 2\tabularnewline
\cline{1-3} \cline{5-7} \cline{9-11} \cline{13-15} 
4 & 3 & 18 &  & 11 & 13 & 10 &  & 18 & 24 & 4 &  & 25 & 31 & 2\tabularnewline
\cline{1-3} \cline{5-7} \cline{9-11} \cline{13-15} 
5 & 4 & 16 &  & 12 & 14 & 8 &  & 19 & 25 & 4 &  & 26 & 32 & 1\tabularnewline
\cline{1-3} \cline{5-7} \cline{9-11} \cline{13-15} 
6 & 5 & 16 &  & 13 & 15 & 8 &  & 20 & 26 & 4 &  & \multicolumn{1}{c}{} &  & \tabularnewline
\cline{1-3} \cline{5-7} \cline{9-11} 
7 & 6 & 16 &  & 14 & 16 & 8 &  & 21 & 27 & 2 &  & \multicolumn{1}{c}{} &  & \tabularnewline
\cline{1-3} \cline{5-7} \cline{9-11} 
\end{tabular}}
\par\end{centering}

\caption{Code distances of the codes of length $32$
used for the construction of concatenated polar codes\label{tab:palette}}
\end{table}

\begin{figure}[h!]
\begin{centering}
\includegraphics[scale=0.7]{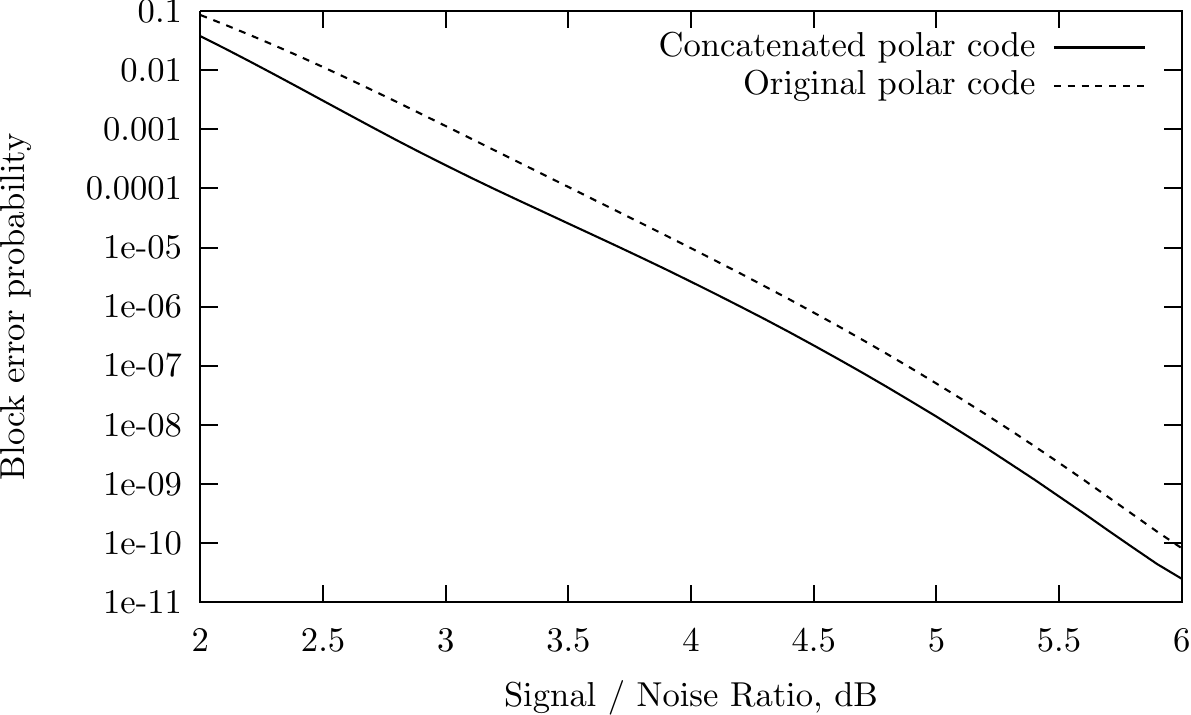}
\par\end{centering}

\caption{Comparison of performance of usual polar code and concatenated polar code
of length $1024$ and rate $\frac12$ on an AWGN channel\label{fig:1024-ecc-vs-simple}}
\end{figure}

\pagebreak{}

\begin{figure}[h!]
\begin{centering}
\includegraphics[scale=0.7]{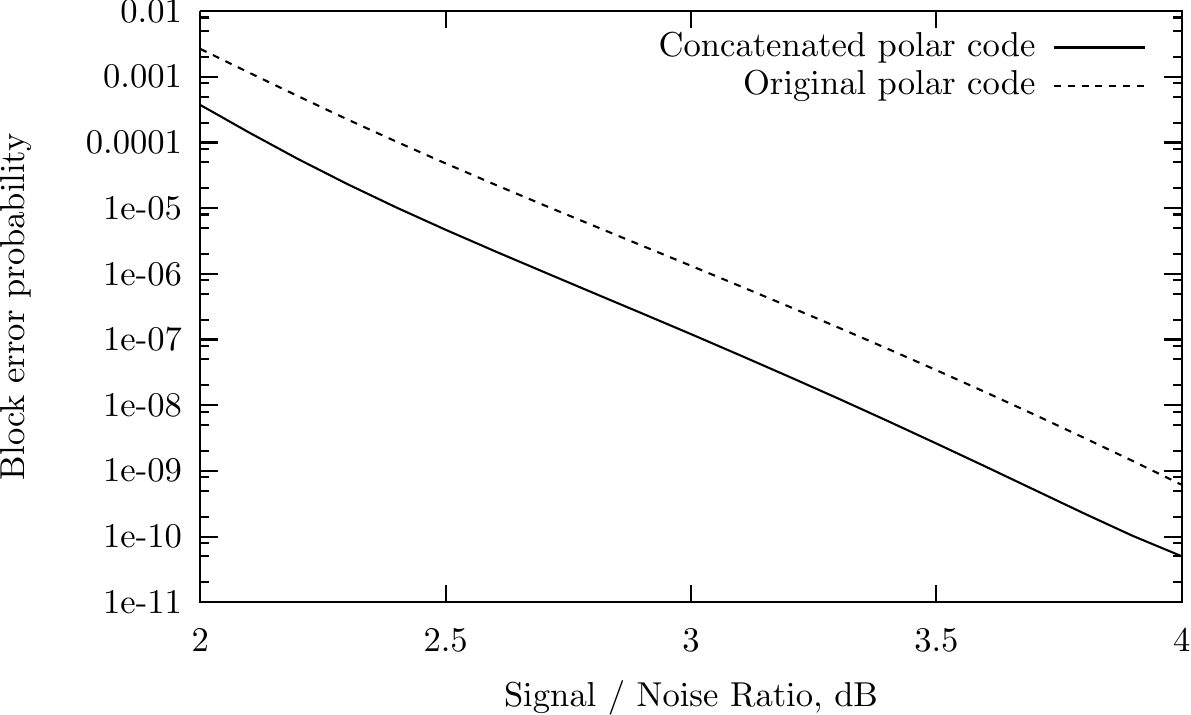}
\par\end{centering}
\caption{Comparison of performance of usual polar code and concatenated polar code
of length $8192$ and rate $\frac12$ on an AWGN channel\label{fig:8k-ecc-vs-simple}}
\end{figure}

\begin{figure}[h!]
\begin{centering}
\includegraphics[scale=0.7]{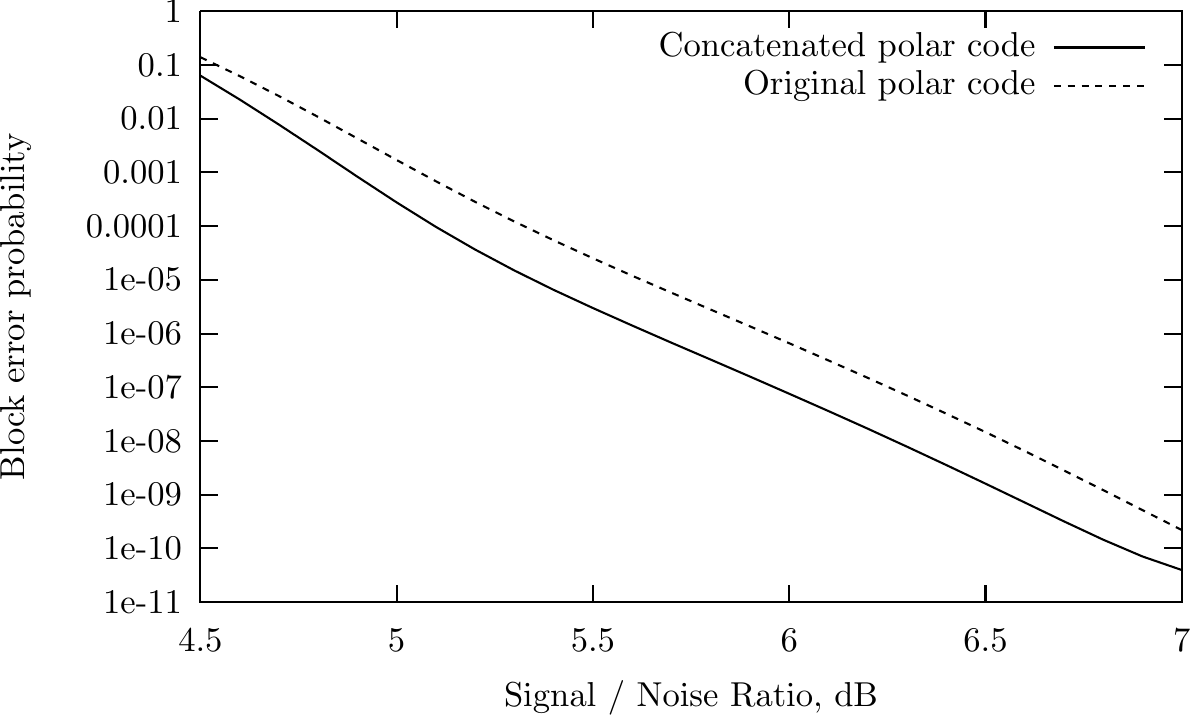}
\par\end{centering}
\caption{Comparison of performance of usual polar code and concatenated polar code
of length $8192$ and rate $\frac34$ on an AWGN channel\label{fig:8k-r75-ecc-vs-simple}}
\end{figure}

\pagebreak{}

It is also interesting to compare the performance of polar code and of concatenated
polar code of the same length and rate. Fig.~\ref{fig:1024-ecc-vs-simple} shows
the performance graph of the concatenated polar code of length
$1024$ and rate $\frac{1}{2}$ which was already presented above
together with the polar code of the same length and rate.
Both codes were constructed for an AWGN channel with SNR$=2.5$dB.
One can see that the concatenated code outperforms the usual one
by an order of magnitude.
The figures~\ref{fig:8k-ecc-vs-simple} and \ref{fig:8k-r75-ecc-vs-simple}
show analogous comparative graphs for the codes of length
 $8192$ and rates $\frac{1}{2}$ and $\frac{3}{4}$, respectively.
Similar to the previous example, concatenated polar codes also outperform the usual
ones approximately by an order of magnitude.


\end{document}